\documentclass[acmsmall]{acmart}

\acmJournal{PACMPL}
\acmVolume{1}
\acmNumber{CONF} 
\acmArticle{1}
\acmYear{2018}
\acmMonth{1}
\acmDOI{} 
\startPage{1}

\setcopyright{none}
\usepackage{booktabs}
\usepackage{url}
\usepackage{subcaption}
\usepackage{xspace}
\usepackage{listings,lstautogobble}
\usepackage{tikz}
\usepackage{caption}
\usepackage{pict2e}
\usepackage{enumitem}
\usepackage{multirow}
\usepackage{adjustbox}
\usepackage{mathtools}	
\usepackage{mdframed}
\usepackage{syntax}
\usepackage{xcolor}
\usepackage[skins]{tcolorbox}
\usepackage{array}
\usepackage{amsmath}
\usepackage{pgfplots}
\pgfplotsset{
	compat=1.4,
	small,
	legend style={
		at={(0.99,0.99)},
		anchor=north east,
		font=\bfseries,
	},
	label style={font=\sffamily\small\bfseries}
}
\usepackage[normalem]{ulem}
\let\UnderScore_

\newcommand\redout{\bgroup\markoverwith
	{\textcolor{red}{\rule[.45ex]{1.5pt}{1.pt}}}\ULon}

\global\mdfdefinestyle{default}{%
	usetwoside=false,%
	linecolor=black,linewidth=1pt,%
	innerleftmargin=5pt,innerrightmargin=5,%
	everyline=true
}

\newcommand{\tabincell}[2]{\begin{tabular}{@{}#1@{}}#2\end{tabular}}

\theoremstyle{definition}
\newtheorem{definition}{Definition}[section]

\makeatletter
\newenvironment{btHighlight}[1][]
{\begingroup\tikzset{bt@Highlight@par/.style={#1}}\begin{lrbox}{\@tempboxa}}
	{\end{lrbox}\bt@HL@box[bt@Highlight@par]{\@tempboxa}\endgroup}

\newcommand\btHL[1][]{%
	\begin{btHighlight}[#1]\bgroup\aftergroup\bt@HL@endenv%
	}
	\def\bt@HL@endenv{%
	\end{btHighlight}%
	\egroup
}
\newcommand{\bt@HL@box}[2][]{%
	\tikz[#1]{%
		\pgfpathrectangle{\pgfpoint{1pt}{0pt}}{\pgfpoint{\wd #2}{\ht #2}}%
		\pgfusepath{use as bounding box}%
		\node[anchor=base west, fill=orange!25,outer sep=.5pt,inner xsep=0.5pt, inner ysep=0.15pt, rounded corners=1pt, minimum height=\ht\strutbox-.1pt,#1]{\raisebox{.01pt}{\strut}\strut\usebox{#2}};
	}%
}
\makeatother

\lstdefinestyle{mystyle}{
	frame=single,
	framexleftmargin=0pt,
	commentstyle=\color{ForestGreen},
	keywordstyle=\color{blue}\bfseries,
	numberstyle=\tiny\color{gray},
	stringstyle=\color{purple},
	basicstyle=\tiny\ttfamily\bfseries,
	breakatwhitespace=false,         
	breaklines=false,                 
	captionpos=b,                    
	keepspaces=true,     
	numbers=none,                    
	numbersep=4pt,                  
	showspaces=false,                
	showstringspaces=false,
	showtabs=false,                  
	tabsize=2,
	language=Java,
	escapechar=\%,
	moredelim=**[is][\btHL]{`}{`},
	moredelim=**[is][{\btHL[fill=red!40]}]{@}{@},
    moredelim=[is][\underbar]{(*}{*)},
		moredelim=[is][\color{black}]{|-}{-|},
		moredelim=[is][\color{blue}\underbar]{|*}{*|},
}
\definecolor{light-gray}{gray}{0.80}
\definecolor{ForestGreen}{RGB}{63,147,88}

\DeclareMathOperator*{\argmax}{argmax}

\usetikzlibrary{pgfplots.groupplots}
\definecolor{RYB1}{RGB}{141, 211, 199}
\definecolor{RYB2}{RGB}{255, 255, 179}
\definecolor{RYB3}{RGB}{190, 186, 218}
\definecolor{RYB4}{RGB}{251, 128, 114}
\definecolor{RYB5}{RGB}{128, 177, 211}
\definecolor{RYB6}{RGB}{253, 180, 98}
\definecolor{RYB7}{RGB}{179, 222, 105}
\pgfplotscreateplotcyclelist{colorbrewer-RYB}{
	{RYB1!50!black,fill=RYB1},
	{RYB2!50!black,fill=RYB2},
	{RYB3!50!black,fill=RYB3},
	{RYB4!50!black,fill=RYB4},
	{RYB5!50!black,fill=RYB5},
	{RYB6!50!black,fill=RYB6},
	{RYB7!50!black,fill=RYB7},
}

\usetikzlibrary{matrix}
\newcommand{\textmtte}[1]{{\fontsize{9}{11}\fontfamily{txtt}\selectfont #1}}
\renewcommand{\texttt}[1]{\textmtte{#1}}
\usepackage{makecell}

\usepackage{multicol}

\newcommand*\circled[1]{\tikz[baseline=(char.base)]{
            \node[shape=circle,draw,inner sep=0.1pt] (char) {#1};}}

\usepackage{xr}
\usepackage{algorithm}
\usepackage{algpseudocode}
\renewcommand*\Call[2]{\textproc{#1}(#2)}
\algnewcommand{\LineComment}[1]{\Statex #1}
\algrenewcommand\algorithmicindent{1.0em}

\usepackage{pifont}

\usepackage{wrapfig}

\usepackage{arydshln}

\usepackage[pscoord]{eso-pic}
\newcommand{\placetextbox}[3]{
  \setbox0=\hbox{#3}
  \AddToShipoutPictureFG*{
    \put(\LenToUnit{#1\paperwidth},\LenToUnit{#2\paperheight}){\vtop{{\null}\makebox[0pt][c]{#3}}}%
  }%
}%

\newcommand{\eg}{\hbox{\emph{e.g.,}}\xspace}
\newcommand{\ie}{\hbox{\emph{i.e.,}}\xspace}
\newcommand{\st}{\hbox{\emph{s.t.}}\xspace}
\newcommand{\wrt}{\hbox{\emph{w.r.t.}}\xspace}
\newcommand{\etc}{\hbox{\emph{etc.}}\xspace}

\newcommand{\tool}{\textsf{Infrared}\xspace}
\newcommand{\met}{\textsf{MBD}\xspace}
\newcommand{\eradicate}{{Eradicate}\xspace}

\begin{document}
\title[Infrared: A Meta Bug Detector]{Infrared: A Meta Bug Detector}	

\author{Chi Zhang}
\affiliation{
	\department{State Key Laboratory for Novel Software Technology Department of Computer Science and Technology}              
	\institution{Nanjing University}            
	\city{Nanjing}
	\state{Jiangsu}
	\postcode{210023}
	\country{China}                    
}
\email{zhangchi\_seg@smail.nju.edu.cn}

\author{Yu Wang}
\affiliation{
	\department{State Key Laboratory for Novel Software Technology Department of Computer Science and Technology}              
	\institution{Nanjing University}            
	\city{Nanjing}
	\state{Jiangsu}
	\postcode{210023}
	\country{China}                    
}
\email{yuwang\_cs@smail.nju.edu.cn}

\author{Linzhang Wang}
\affiliation{
	\department{State Key Laboratory for Novel Software Technology Department of Computer Science and Technology}              
	\institution{Nanjing University}            
	\city{Nanjing}
	\state{Jiangsu}
	\postcode{210023}
	\country{China}                    
}
\email{lzwang@nju.edu.cn}          

\begin{abstract}


The recent breakthroughs in deep learning methods have sparked a wave of interest in learning-based bug detectors. Compared to the traditional static analysis tools, these bug detectors are directly learned from data, thus, easier to create. On the other hand, they are difficult to train, requiring a large amount of data which is not readily available. In this paper, we propose a new approach, called \textit{meta bug detection}, which offers three crucial advantages over existing learning-based bug detectors: bug-type generic (\ie capable of catching the types of bugs that are \textit{totally unobserved} during training), self-explainable (\ie capable of explaining its own prediction without any external interpretability methods) and sample efficient (\ie requiring substantially less training data than standard bug detectors). Our extensive evaluation shows our meta bug detector (\met) is effective in catching a variety of bugs including null pointer dereference, array index out-of-bound, file handle leak, and even data races in concurrent programs; in the process \met also significantly outperforms several noteworthy baselines including Facebook Infer, a prominent static analysis tool, and FICS, the latest anomaly detection method.


\end{abstract}
\keywords{Meta Bug Detection, Graph Neural Networks}
\maketitle
	
\section{Introduction}


The recent breakthroughs in deep learning have created a new opportunity for static bug finding which have been dominated by formal methods~\cite{Clarke332656,Cousot512973,King1975}.
This new strand of bug detection work~\cite{allamanis2018learning,pradel2018deepbugs} is motivated by the observation that bug patterns exist~\cite{Hovemeyer1052895}, thus, models can be trained to capture them. Compared to the traditional symbolic-, logic-based approaches, this data-driven approach helps reduce the complexity of the creation of  static bug finders, which now can be directly learned from data using various machine learning models that are publicly available.
%
However, such learning-based bug detection suffers from their own weaknesses, one in particular, is that they require large datasets of known bugs for training, which can be difficult to assemble. Specifically, manually curated benchmarks~\cite{10.1145/2610384.2628055, saha2018bugs,widyasari2020bugsinpy} are hard to scale, especially for deep neural networks which are notoriously data-hungry. On the other hand, automated methods~\cite{chen2018sequencer,Dinella2020HOPPITY,Wang3428205} that treat code commits as sources of bugs hurt the quality of training data. To start, code commits are not necessarily bug fixes (\eg refactoring, optimization, \etc), even if they are, they may not be the correct, principled patches that address the root cause of bugs, instead, they can be hacks, work-arounds that developers are known to use to make code pass test cases. Therefore, how to accurately identify the context of a bug given the location of the fix remains a significant challenge. Recently, a new bug mining method~\cite{Madeiral2019,BugSwarm-ICSE19} has been proposed that relies on modern continuous-integration approaches, like Travis-CI. As the technology is still at its early stage, it has not grown datasets to the scale needed to train deep neural networks. Finally, there are also works that resort to synthetic bugs for the sufficiency of training data~\cite{ferenc2018public,ferenc2020automatically}, considering synthetic bugs are unlikely to fit to the distribution that describes the real bugs, this approach is also flawed. 

In addition to the lack of large-scale, high-quality datasets, existing learning-based bug detectors have another significant limitation: they can only learn patterns specific to the type of bugs that are given as training data, in
other words, in test-time they would simply surrender to a type of bugs that were unobserved during training.

\noindent
\textbf{\textit{A Meta Bug Detector (\met).}}\, To address the weaknesses of the aforementioned prior works, this paper proposes a new approach to learning bug detectors,
called \textit{meta bug detection}.
Although meta bug detector (\met) is also a learning-based bug detector to its core,
it is built upon a fundamentally different concept:
instead of learning the patterns specific to each bug type,
the approach taken by almost all existing bug detectors, \met opts to learn how bugs make programs inconsistent with the norm independently of the bug types --- hence the name \textit{Meta}. Under this guiding principle, \met offers many benefits.

\begin{figure*}[h!]
\newcommand\x{2pt}
\newcommand\y{10pt}
	\centering
	\captionsetup{skip=2pt}
    \begin{multicols}{2}
     \centering

	\begin{subfigure}{0.48\textwidth}
	\setlength{\abovecaptionskip}{\x}
	\setlength{\belowcaptionskip}{\y}
		\lstset{style=mystyle}
\begin{lstlisting}[morekeywords={Object, String}, basicstyle=\scriptsize\ttfamily\bfseries]
public boolean equals(Object obj) {
  Match o = (Match) obj;...
  if (o.left != null && o.right != null) {...
    this.right = @this.right.strip()@;
  }
  if (... && @this.right != null@) {
    return ... && this.right.equals(o.right);
  } 
}
\end{lstlisting}
		\caption{A null pointer dereference.}
		\label{fig:cincbug}
	\end{subfigure}
      \par
	\begin{subfigure}{0.48\textwidth} 
	\setlength{\abovecaptionskip}{\x}
	\setlength{\belowcaptionskip}{\y}
		\lstset{style=mystyle}
\begin{lstlisting}[morekeywords={Object, String},basicstyle=\scriptsize\ttfamily\bfseries]
protected void processMethod(Object bean, 
    String beanName, Method method) {
  ApolloJsonValue apolloJsonValue = 
    AnnotationUtils.getAnnotation(method,...);

  if (@apolloJsonValue == null@) 
    return;

  String placeholder = 
      @apolloJsonValue.value();@

}
\end{lstlisting}
		\caption{A correct example \wrt~(\subref{fig:cincbug}).}
		\label{fig:cinccorrect1}
	\end{subfigure}
      \par
	\begin{subfigure}{0.48\textwidth} 
	\setlength{\abovecaptionskip}{\x}
		\lstset{style=mystyle}
\begin{lstlisting}[morekeywords={Object, String},basicstyle=\scriptsize\ttfamily\bfseries]
private void checkTypeLiteral(JCTree node, 
    Symbol sym) {
  if (@sym == null@ || 
      sym.kind != Kinds.Kind.TYP) 
    return;
  Path jarPath = 
    getJarPath(@sym.enclClass()@, 
    platformJars);
  if (jarPath != null) {
    ...}
}
\end{lstlisting}
		\caption{Another correct example \wrt~(\subref{fig:cincbug}).}
		\label{fig:cinccorrect2}
	\end{subfigure}
      \par
	\begin{subfigure}{0.48\textwidth}
	\setlength{\abovecaptionskip}{\x}
	\setlength{\belowcaptionskip}{\y}
	 \lstset{style=mystyle}
\begin{lstlisting}[morekeywords={Object, String}, basicstyle=\scriptsize\ttfamily\bfseries]
public InputStream openFile() {
  final InputStream tmp = zipFile.openFile();
  if (tmp != null) try {
      final ZipInputStream zis = 
          new ZipInputStream(tmp);
      ZipEntry ze = zis.getNextEntry(); ...
      @zis.close();@
    } catch (IOException e) {} ...
}
\end{lstlisting}
	 \caption{A file handle leak.}
	 \label{fig:crsbug}
	\end{subfigure}
      \par	 
	\begin{subfigure}{0.48\textwidth}
	\setlength{\abovecaptionskip}{\x}
	\setlength{\belowcaptionskip}{\y}
	 \lstset{style=mystyle}
\begin{lstlisting}[morekeywords={Object, String}, basicstyle=\scriptsize\ttfamily\bfseries]
void persistLocalCacheFile(File baseDir) {
  File file = assembleLocalCacheFile(...);
  OutputStream out = null;
  try {
    out = new FileOutputStream(file);
    ...
  } catch (IOException ex) {} finally {
    ...
    try {
      @out.close();@
    } catch (IOException ex) {}
  }}
\end{lstlisting} 
	 \caption{A correct example \wrt~(\subref{fig:crsbug}).}
 	 \label{fig:crscorrect1}
	\end{subfigure}
	      \par
	 \begin{subfigure}{0.48\textwidth}
	\setlength{\abovecaptionskip}{\x}
	 \lstset{style=mystyle}
\begin{lstlisting}[morekeywords={Object, String}, basicstyle=\scriptsize\ttfamily\bfseries]
public void writeFile(File inputFile, 
    String jarPath) throws IOException {
  FileInputStream fis = 
      new FileInputStream(inputFile);
  try {
    JarEntry entry = new JarEntry(jarPath);
    entry.setTime(inputFile.lastModified());
    writeEntry(fis, entry);
  } catch (IOException ex) {...}finally {
    @fis.close();@
  }}
\end{lstlisting}
	 \caption{Another correct example \wrt~(\subref{fig:crsbug}).}
	 \label{fig:crscorrect2} 
	\end{subfigure}
    \end{multicols}

	\caption{\met's perspective in catching null pointer dereference and file handle leak. The bug in (\subref{fig:cincbug}) is caused by the wrong order of null check and dereferencing. The bug in (\subref{fig:crsbug}) is triggered when code in  \texttt{try} clause throws an exception, in which case, \texttt{catch} clause immediately get executed causing \texttt{zis.close()} to be skipped, hence the leak.}	
	\label{fig:motivatingExample}
	\vspace{-5pt}
\end{figure*}

First and foremost, it reduces the difficulty level of the learning task. This is because compared to the bug patterns per se, the manners in which buggy programs contradict correct programs are more obvious and far easier to learn. We illustrate this point using the programs in Figure~\ref{fig:motivatingExample}. For the well-known null pointer dereference bugs, a standard bug detector would learn to recognize a program execution on which a pointer first gets assigned the value null and then dereferenced later. This is no easy task considering an accurate bug detector must precisely capture the pattern without raising a large number of false positives. Specifically for the bug in Figure~\ref{fig:cincbug}, there is no clear signal that indicates this pattern indeed exists in the program,
however, if we contrast the buggy program to the correct code in Figure~\ref{fig:cinccorrect1} and~\ref{fig:cinccorrect2}, the inconsistency emerges, that is, in the correct code, dereferencing a pointer happens after the very pointer is checked against the null value whereas the two events happen in a reverse order in the buggy code. For this reason, the program in Figure~\ref{fig:cincbug} is ranked by \met as one of the most probable anomalies to those that correctly handled potential null pointers. A direct benefit that \met enjoys from performing an easier learning task is the substantial reduction of training data. Our experiment shows a few thousand programs {suffice} to train a \met with decent performance. This is in fact a crucial advantage of \met's over the existing learning-based bug detectors especially considering that large-scale, high-quality, real-world defect datasets are not available. Given this constraint, \met's approach of learning inconsistency between buggy and correct programs promises a pathway forward. 

Second, \met is capable of catching the type of bugs it has never encountered during training, a property that existing learning-based bug detectors do not possess. Because, as we explained, those bug detectors can only deal with the bug types they are trained on and there are little to no commonalities across distinct bug types. We show, however, the commonalities indeed exist from \met's perspective and how \met exploits this phenomenon to detect the unseen type of bugs in test-time. Take the three programs in Figure~\ref{fig:crsbug}-\ref{fig:crscorrect2} for example. From a standard bug detector's perspective, detecting the pattern of file handle leak --- a handle to a file is requested but not freed when it is no longer used --- is nothing like detecting the pattern of null pointer dereference. However, not only {does} \met detect the program in Figure~\ref{fig:crsbug} as a contradiction to those in Figure~\ref{fig:crscorrect1} and~\ref{fig:crscorrect2} but does so by learning exclusively from examples prepared for null pointer dereference like those in Figure~\ref{fig:cincbug}-\ref{fig:cinccorrect2}. The key to \met's success is that \met detects the manner in which buggy code contradict correct code in the file handle leak is similar to that in the null pointer dereference. Specifically, a statement is misplaced in the buggy code \wrt to where the statement is in the correct code. Regarding the file handle leak, it's the stream closing operation which should have been placed in the \texttt{finally} clause rather than the \texttt{try} clause, whereas for the null pointer dereference, the dereferencing should have occurred after the null check as we explained previously.

\begin{figure*}[t!]
	\centering
	\captionsetup{skip=4pt}
	
	\parbox{.51\linewidth}{
		\lstset{style=mystyle}
\begin{lstlisting}[morekeywords={Object, String},basicstyle=\scriptsize\ttfamily\bfseries]
public class Match {
  private final String left;
  private final String right;
  public Match(String left, String right) {
    this.left = left;
    this.right = right;
  }

  public boolean equals(Object obj) {
    Match o = (Match) obj;
    if (o == null) return false;
    if (o.left != null && o.right != null) {
      ...
      this.left = this.left.strip();
3:    this.right = this.right.strip();
    } 
    if (this.left != null && this.right != null)
      ...
  }
}

void Main() {
1:Match X = new Match("left", null);
  Match Y = new Match("left", "right");
  ...
2:X.equals(Y);
}
\end{lstlisting}
	}
	\caption{The trace format in which \met explains its prediction (using Figure~\ref{fig:cincbug} as an example): (1) underlined statements are the key steps of the trace; (2) the number marks the order in which statements are executed; (3) the statement within the bounding box is the program point at which the bug occurs.}
	\label{fig:explan}
	
	\vspace{-6pt}
\end{figure*}

Third, \met addresses a key issue for which existing learning-based bug detectors have long been lamented. That is, apart from the prediction results, they can not provide concrete feedback upon which developers can act. This shortcoming casts a doubt on the practical utility of learning-based bug detectors. On the contrary, \met is capable of explaining its own predictions, in particular, it follows a standard practice from static analysis tools for reporting bugs to end users. That is, for a program predicted as buggy, \met outputs a buggy trace --- a program path
along which the bug can be triggered --- to illustrate the bug. Our insight is to use inconsistencies between code snippets, the reason that \met considers programs to be buggy, to produce the buggy trace. By means of global attention mechanism (which we will present later), we first capture the part of the buggy program \met deems to be inconsistent with the correct code, and then reduce it into a program trace.
For example, Figure~\ref{fig:explan} shows the buggy trace for \met's prediction for the program in Figure~\ref{fig:cincbug}. In light of this new explainability notion, we redefine \met's output for a buggy program to be the buggy trace, which is considered to be correct only if the buggy trace triggers the bug. This way \met will behave like a static analysis tool when it comes to bug reporting, ensuring its usefulness in practice. 
%

We note that the key concept that underlies the design of \met is not new. 
Prior works took this approach to detecting errors in system code~\cite{Engler502041}, incorrect API usages~\cite{197149,9519443} and other specific types of bugs~\cite{10.1145/581339.581377,1883978.1883982,7886915}. However, \met surpasses these works in many ways. First, it is capable of catching the type of bugs it has not encountered during training.
On the contrary, anomaly detection works cannot deal with the classes of anomaly that their design was not modeled after. Second, it does not require human experts to define rules or heuristics specific to the type of bugs it aims to catch. Third, it learns from both buggy and correct programs, enabling the machine learning model to accurately distinguish
the two classes of examples. In contrast, anomaly detection approaches often infer information from existing code by learning only from correct programs, hence, 
they usually suffer from a high false positive rate.
%

%
%
%
%
%
\placetextbox{0.452}{0.7335}{\setlength{\fboxsep}{3.5pt}\fbox{\qquad\qquad\qquad\qquad\qquad\qquad\qquad\quad}}
\placetextbox{0.457}{0.6425}{\underbar{\qquad\qquad\qquad\qquad\qquad\qquad\qquad\;}}
\placetextbox{0.362}{0.6093}{\underbar{\qquad\qquad\;\;\;\,}}

We face two technical challenges to realize \met: (1) developing a machine learning model that can accurately capture the semantic patterns that programs denote, and does so in a transparent manner to facilitate its explainability efforts;
(2) formulating \met's learning objective to accomplish its mission of being a bug-type generic detector. To resolve the first challenge, we utilize Graph Neural Network (GNN) for its supreme performance achieved in a wide range of programming-language tasks. Furthermore, we propose a global attention mechanism in which GNN learns to focus on nodes that express the essential characteristics of the graph and de-emphasize the rest. Note that this method differs from the widely used graph attention mechanism (\eg graph attention network~\cite{velikovi2018graph}) in which nodes are prioritized within a neighborhood as opposed to the entire graph. In the meta bug detection setting, our global attention mechanism has shown to significantly improve GNN's learning precision. In addition, it also provides an insight {into} GNN's inner workings, laying the foundation of the aforementioned \met's self-explainability. As for the second challenge, we first prepare a dataset that manifests the common ways that buggy code is inconsistent with correct code. Then, for each type of inconsistency, we collect buggy
programs (in smaller quantities deemed as the anomalies) and correct programs (in larger quantities forming the main cluster) in a way that every buggy program is inconsistent with every correct program in the same manner. 
Obviously, directly fitting \met to the specific patterns that buggy programs display will not make it generalize to bugs beyond this dataset, instead, we train \met to recognize the patterns of inconsistency between buggy and correct code. Technically, we define a loss function based on triplet loss~\cite{Matthew03} with which \met learns to embed buggy programs further away from the main cluster of correct programs as anomalies for each type of the inconsistency. Because the inconsistencies from which \met learns can also describe other bugs outside of this dataset,
our training regime 
enables \met to catch the type of bugs that are unobserved during 
training.



We have realized \met as a new bug detector, \tool, and extensively evaluated it. First, we show, on 99 bugs we  extracted from existing datasets, \tool is effective, accurately separating bugs from correct programs without being exposed to the types of bugs during training. We further evaluate the effectiveness of \tool on Defects4J ~\cite{10.1145/2610384.2628055}, a well-established bug dataset. Regarding the baseline of this experiment, we pick the two most effective static analysis tools --- \eradicate~\cite{Eradicate} and Infer~\cite{10.1145/2049697.2049700} --- according to the results of an extensive study recently conducted by~\citet{Tomassi2021}. We also include FICS~\cite{263838}, arguably the state-of-the-art anomaly detection work. When trained on 1091 programs, \tool outperforms all baselines by a comfortable margin. Finally, we evaluate \tool in detecting data race, the most reliable indicator of errors in concurrency. We choose data race because we believe it presents a bigger challenge than the previous tasks due to the unique characteristics that concurrent programs exhibit. Remarkably, using the same 1091 training instances without a single concurrent program, \tool beats out RacerD~\cite{10.1145/3276514}, one of the most recent advances in static race detection. 


\vspace{3pt}
\noindent
\textbf{\textit{Contributions.}}\, We make the following main contributions:

\begin{itemize}
	\item We introduce the concept of \met for training bug type-generic, self-explainable, and sample efficient bug detectors.
	\item We define a loss function with which \met learns to recognize how buggy code are inconsistent with the correct code in a bug type-generic fashion.	
	\item We propose global graph attention which not only improves GNN's learning precision but also provides an insight {into} GNN's inner workings, effectively moving GNN a step closer to being an explainable model.
	\item We report our extensive evaluation of \tool, a realization of \met, on three benchmarks. Results show \tool is effective, consistently outperforming several prominent baselines.

\end{itemize}

\vspace{3pt}
\noindent
\textbf{\textit{Paper Organization.}}\,
The rest of the paper is structured as follows. Section~\ref{sec:over} illustrates \met's workflow. Section~\ref{sec:gl} gives a detailed presentation on \met's learning approach. Next, we describe our extensive evaluation of \tool, an implementation of \met (Section~\ref{sec:eva}). Finally, we survey related work (Section~\ref{sec:rela}) and conclude (Section~\ref{sec:conc}).

\section{Overview}
\label{sec:over}

\begin{figure}[tb!]
	\centering
	\begin{subfigure}{1\textwidth}
		\centering
		\includegraphics[width=\textwidth]{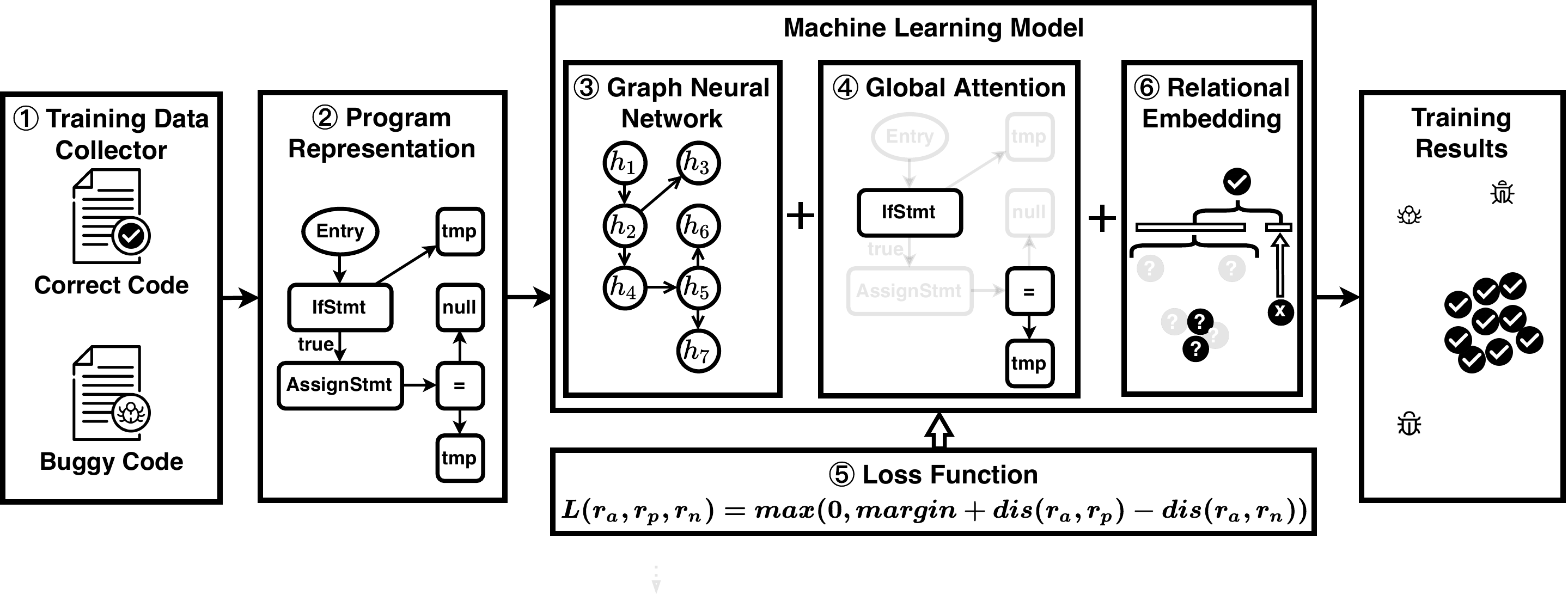}
		\vspace*{-30pt}
		\caption{Training.
		}
		\label{fig:training}
	\end{subfigure}
	
	\vspace{15pt}
	
	\begin{subfigure}{1\textwidth}
		\centering
		\includegraphics[width=\textwidth]{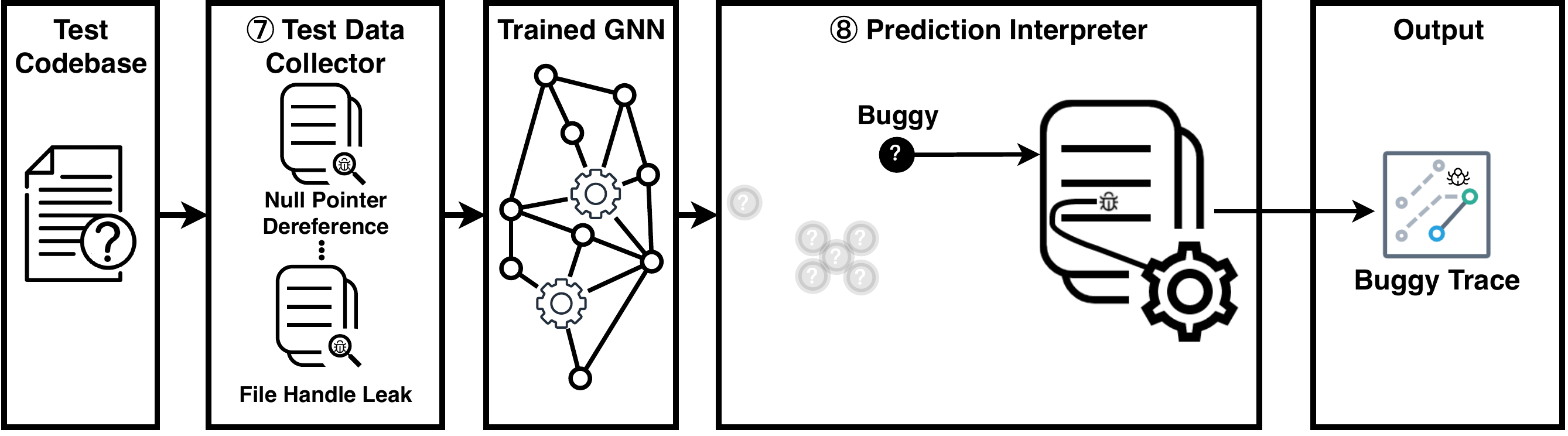}
		\caption{Test.}
		\label{fig:testing}
	\end{subfigure} 	
	\caption{\met's workflow.}
	\label{fig:keycom}
	\vspace{-5pt}
\end{figure}

In this section, we present a high-level workflow of \met, in particular, we introduce \met's components for both training and test. 

\vspace{3pt}
\noindent
\textbf{\circled{1}} \textbf{\textit{Training Data Collector}} Curating a quality dataset lays the foundation for effective \met. The primary criteria we define for the dataset is the inclusion of a diverse set of inconsistent code (\eg missing or misplacing checks, using wrong variables/operators/APIs, \etc). By inconsistency code, we mean a set of programs containing a few bugs and far more correct programs, among them every bug is inconsistent with every correct program.\footnote{we include far more correct programs than bugs to reflect the ratio between buggy and correct programs in any mature codebase in the real-world.}
We first attempt to use existing tools~\cite{197149,263838} on inconsistency code detection to collect data. Specifically, we set them up to run on older versions of several open-source projects that are known to be buggy. After a manual inspection, we find the results of those tools are exceedingly noisy as existing tools routinely miss the known bugs; even for the few they manage to catch, the respective correct code do not show any inconsistencies.

Next, we attempt to construct our dataset based on existing datasets of bugs. Although those datasets include detailed information for a bug (\eg location of the bug, patches of the bug, the codebase the bug is in, \etc), they do not provide a compact code snippet to describe the bug, instead, they specify the version of the codebase in which the bug is found. Considering machine learning models are most effective when given accurate data to learn, we must identify from the whole program (\ie the entire codebase) only the code related to the bug. In addition, finding the corresponding correct programs after the (more precise representation of the) bug is obtained is another challenge we must overcome.
We build a fully automated pipeline to resolve both challenges. First, given the version of a codebase that contains a bug, we compute a slice~\cite{WeiserSlice} from the whole program using the statement where the bug occurs (which is provided by existing datasets) as the criterion. Since our slice contains all the dependencies related to the bug, it is a safe representation of the bug for our inconsistent code. Next, we patch an extracted bug slice to obtain its fixed version, which we use to search for correct programs that are inconsistent with the bug. The reason we can not directly take the fixed version as the correct program is to prevent a distribution discrepancy between training and test because bug detectors will never be asked to distinguish bugs from their fixed versions in test-time. For convenience, the search of the correct code happens within the codebase of the bug, in particular, we collect syntactically similar programs (based on the tree-edit distance between abstract syntax trees) to the fixed version of the bug. This is a reasonable design choice because the semantic inconsistency already exists between the bug and its fixed version, thus, programs that are syntactically similar to the fixed version are likely to maintain the semantic inconsistency. Finally, for efficiency purposes, we aggregate inconsistent code that share a large number of correct programs, a strong indicator that those inconsistent code express the same kind of inconsistency. Inevitably, our automated pipeline would induce noise into our dataset as not every correct program is guaranteed to be inconsistent with every bug in the inconsistent code. However, this is acceptable because \met will not get to work with perfectly inconsistent code during test either. If \met is pressured to learn to be tolerant on data with some noise, it will better cope with the noisy data it sees in the wild. 



\begin{figure}[t]
	\centering
	\includegraphics[width=\textwidth]{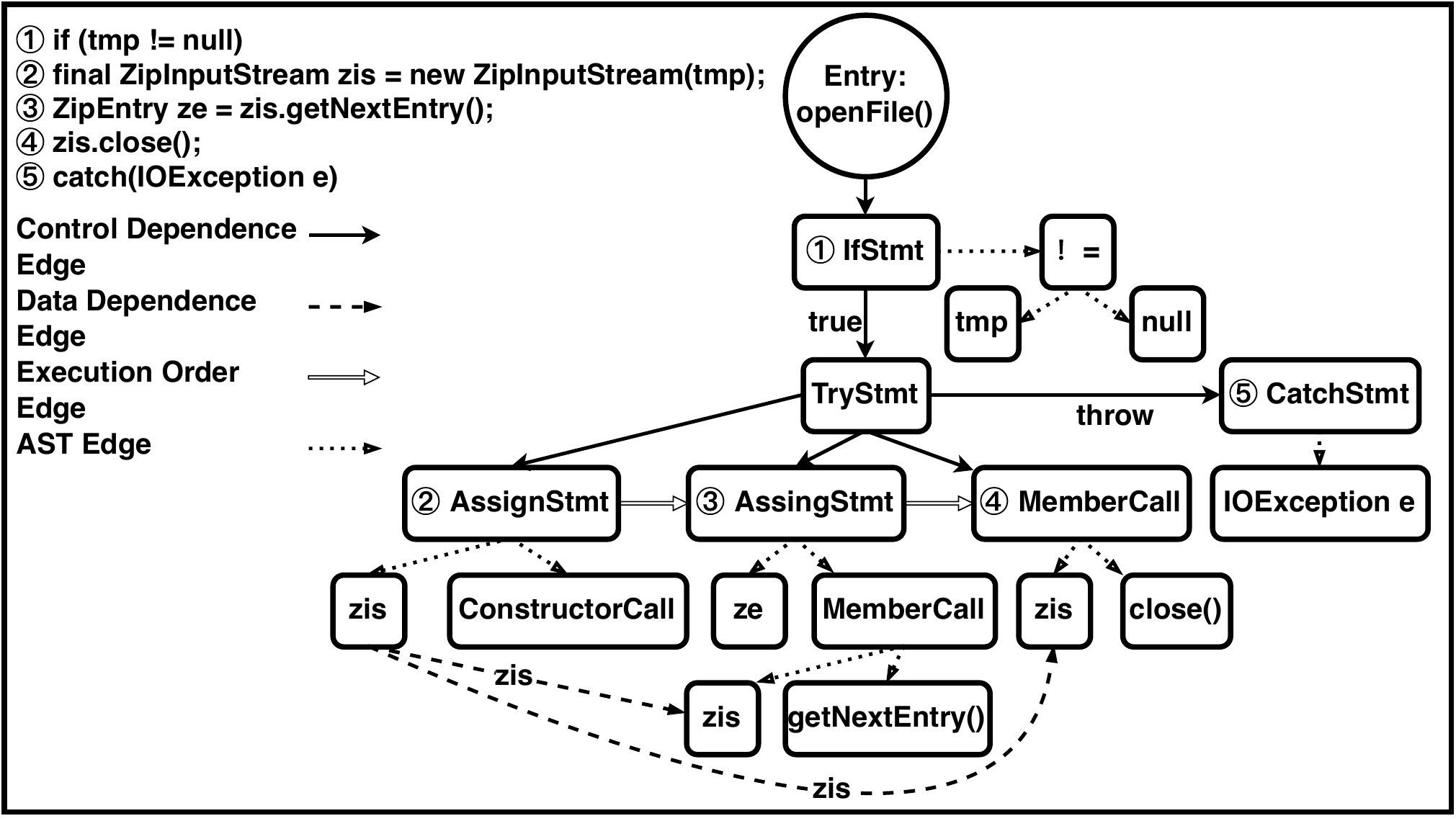}
	\caption{Representing the program in Figure~\ref{fig:crsbug}. We have omitted part of the program for simplicity and brevity. For each numbered root node, we show the corresponding statement in the top-left corner.}
	\label{fig:pdg}
	\vspace*{-5pt}
\end{figure}

\vspace{3pt}
\noindent
\textbf{\circled{2}} \textbf{\textit{Program Representation}} (Section~\ref{subsec:graph}).\, We represent programs using program dependence graph (PDG), 
a principled, systematic graph representation, which makes explicit both the data and control dependence for each operation in a program. Because PDG does not preserve the execution order among statements that are control dependent on the same statement. 
%
We introduce a new type of edges to link those statements in the order they are executed.
To refine the granularity of our graph representation, we encode each statement --- a node in a standard PDG --- with their abstract syntax trees (AST). Accordingly, we make the following two modifications to our graph. First, we reset the control dependence edges and execution order edges to connect the root nodes of statements' AST. Second, we push down the data dependence edges to connect the leaf nodes of statements' AST. Figure~\ref{fig:pdg} depicts the graph representation of the program in Figure~\ref{fig:crsbug}.

\vspace{3pt}
\noindent
\textbf{\circled{3}}\textbf{\textit{ Graph Neural Network.}}\, For performance consideration, we choose graph neural network (GNN)~\cite{1555942} as the underlying machine learning model for \met, specifically, the category of GNNs that is powered by message-passing mechanism~\cite{10.5555/3305381.3305512}. At a high-level, the goal of GNN is to learn node embeddings based on the graph structure. Such node embeddings are numeric vectors that represent the state of nodes. Technically, GNN approaches this problem in a step-wise manner. In each step, each node first sends messages to its neighboring nodes, then aggregates the messages it received from the neighboring nodes to compute its representation for the next step. Below we use GGNN's~\cite{li2015gated} approach as an illustrative example. Equation~\ref{equ:mes1} computes a message ${m}_n$ for node $n$ through $f(\cdot)$ (\eg a linear function) on the representations of its neighboring nodes $\Omega_n$. Next, a $\mathit{GRU}$ takes ${m}_n$ and $h_{n}$ --- the current representation of node $n$ --- to compute the new state $h'_{n}$ (Equation~\ref{equ:mes2}).

\noindent\begin{minipage}{.5\linewidth}
	\vspace{5.5pt}
	\begin{equation}
		{m}_n = \sum_{\mathclap{u\in \Omega_n}} f(h_{u}) \label{equ:mes1}
	\end{equation}
\end{minipage}
\begin{minipage}{.5\linewidth}
	\vspace{-5.5pt}
	\begin{equation}
		h'_{n} = \mathit{GRU}({m}_n, h_{n})             \label{equ:mes2}
	\end{equation}
\end{minipage}
	

\vspace{3pt}
\noindent
 \textbf{\circled{4}} \textbf{\textit{Global Graph Attention}}
(Section~\ref{subsec:atten}).\,
Despite GNN's cutting-edge performance, they can still suffer from precision issues, especially when dealing with large graphs.
This is mainly because of the high cost of its underlying message-passing procedure. Ideally, every node should pass messages directly or indirectly to every reachable node in the graph to allow sufficient information exchange. However, such an expensive propagation is hard to scale to large graphs without incurring a significant precision loss. To overcome this challenge, we propose a global graph attention mechanism that {prioritizes} nodes in the message-passing procedure. Our intuition is that if GNN can focus on nodes that are essential to expressing the characteristics of graphs, GNN would become more precise in capturing patterns from the graph data. 
To contextualize our intuition in the settings of bug detection, it is clear that a large part of the buggy code is usually irrelevant to the semantics bugs denote, therefore, we can downplay a considerable number of nodes from the graph, which not only helps to reduce the cost of the underlying message-passing process, but more importantly, increases GNN's overall accuracy in recognizing the bug patterns from the few remaining nodes.

Through this global attention mechanism, an attention weight will be allocated to every node in the graph based on its relation with a \textit{meta} node, a new node that we add into the graph for approximating the global properties of the graph. The attention weight then determines the level of influence a node can have on GNN's message-passing procedure. Specifically, nodes at the core of the graph will receive higher attention weights, thus, start dominating how GNN learns from the graph. On the contrary, nodes on the fringe of the graph will receive lower attention weights. As a result, they become marginalized and eventually fade out of the graph. Section~\ref{subsec:atten} presents the details of our global attention mechanism.


\vspace{3pt}
\noindent
 \textbf{\circled{5}} \textbf{\textit{\met's Loss Function}}
(Section~\ref{subsec:loss}).\, 
In order to be bug-generic, \met cannot adopt the approach of existing bug detectors which directly learns the patterns from specific types of bugs. However, since bugs are always inconsistent with the correct programs regardless of their types,
we can train \met to recognize the inconsistency in a bug-type generic manner. 
As described in \textbf{\circled{1}} \textbf{\textit{Training Data Collector}}, every inconsistency code contains far more correct programs than buggy programs.
Therefore, we can train \met to recognize the inconsistency in an inconsistent code by placing every bug further away from the average of all programs than every correct program in the embedding space. Equation \ref{equ:trainsema} gives the mathematical formulation.
\begin{equation}
\label{equ:trainsema}
\forall p\in P\, \forall q \in Q\, \lvert g_\theta(p)-\bar{r} \rvert > \lvert g_\theta(q)-\bar{r} \rvert
\end{equation}
where $g_\theta$ is the GNN with parameter $\theta$; $P$/$Q$ denotes the set of buggy/correct programs for one type of inconsistency, and $\bar{r}$, the average program, is defined as follows.
\begin{equation}
\label{equ:referenceset}
\bar{r} = \frac{\sum_{p \in P} g_\theta(p) + \sum_{q \in Q} g_\theta(q)}{\vert P \cup Q \vert}
\end{equation}
In Section~\ref{subsec:loss}, we discuss how to convert Equation~\ref{equ:trainsema} into \met's loss function based on the triplet loss paradigm~\cite{Matthew03}.

\vspace{3pt}
\noindent
\textbf{\circled{6}} \textbf{\textit{Relational Embedding.}}\, A potential issue with \met's learning approach (formulated by Equation~\ref{equ:trainsema}) is that the embeddings 
are myopic in the sense that each program $ x \in P\cup Q$ gets embedded by $g_\theta$ independently of other programs. Instead, we argue $(P\cup Q)\setminus\{x\}$ should have an influence on how we embed $x$ through $g_\theta$. The reason is that some program $y$ in $(P\cup Q)\setminus\{x\}$ may be particularly indicative of $x$'s correctness such as the program in Figure~\ref{fig:cinccorrect1} is of the program in Figure~\ref{fig:cincbug}, in which case it is beneficial to account $y$ while embedding $x$. Some evidence of this is presented in Section~\ref{subsec:deep}.
For this reason, we propose to embed each program $x$ in $P\cup Q$ through a function which takes as input $(P\cup Q)\setminus\{x\}$ in addition to $x$ (\ie $g_\theta(x)$ becomes $g_\theta(x,(P\cup Q)\setminus\{x\})$). Thus, $(P\cup Q)\setminus\{x\}$ now can affect how we embed $x$. Technically, 
we apply the attention mechanism to compute a weighted sum of the embeddings of programs in $(P\cup Q)\setminus\{x\}$ \wrt $x$ (Equation~\ref{equ:wei}). This way programs that should have played a bigger role in embedding $x$ will receive more attention. 
\begin{align}
	R            &= (P \cup Q) \setminus \{x\} \nonumber \\
	z            &= \sum_{y\in R} \alpha(g_\theta(x), g_\theta(y))g_\theta(y) \label{equ:wei}\\
    \alpha(g_\theta(x), g_\theta(y)) &= \frac{\mathit{exp}(g_\theta(x) \odot g_\theta(y))}{\sum_{w\in R}\mathit{exp}(g_\theta(x) \odot g_\theta(w))} 	\label{equ:aw}
\end{align}

Instead of directly taking $z$ as the relational embedding of $x$, we then perform several steps of reads from a Long-Short Term Memory (LSTM)~\cite{10.1162} in an attempt to add depth to the attention. In principle, this allows the model to increasingly attend to the crucial programs in $R$ and de-emphasize the rest. Technically, we follow~\citet{process2016} from their ``Process'' block  --- a module designed for similar purposes to ours. 
\begin{align}
	\hat{\delta}_{\theta}^{k+1}(x), c^{k+1} &= \mathit{LSTM}(g_\theta(x), [\delta_{\theta}^{k}(x), z^k], c^{k})  \nonumber\\
	\delta_{\theta}^{k+1}(x)                &= \hat{\delta}_{\theta}^{k+1}(x) + g_\theta(x)\label{equ:rela}
\end{align}
where $\delta_{\theta}^{k}(x)$ is the relational embedding of $x$ after k-steps of read, $\delta_{\theta}^{0}(x)$ is initialized to $g_\theta(x)$, $z^0$ is defined in Equation~\ref{equ:wei}. Later, $g_\theta(x)$/$g_\theta(y)$ will be replaced with $\delta_{\theta}^{k}(x)$/$\delta_{\theta}^{k}(y)$ for computing $z^{k}$. $c^0$ is assigned randomly.

\vspace{3pt}
\noindent
\textbf{\circled{7}} \textbf{\textit{Test Data Collector.}}\,
Directly applying a trained \met to find bugs in a test codebase is not the ideal way forward. This is because the notion of inconsistency is valid only if test programs share common properties (\eg inconsistency has no validity concerning programs that trim a string and sort an array). Deploying \met to scan an entire codebase will unavoidably make \met work with totally unrelated code, which clearly violates the premise of \met's approach. Therefore, the first task we need to resolve is to collect the set of test programs to which \met is applicable for identifying the inconsistencies. A simple solution is to group together test programs that could potentially induce the same type of bugs.
Technically, \wrt the nature of each common type of bugs, we develop a data-flow analysis to collect test programs. Considering  the machine learning model (further down \met's workflow) will not be able to bring back any bugs if they are missed by this analysis, we design our analysis to be sound.
Below, we illustrate this analysis using null pointer dereference as an example. We first find out the variables $V$ in the entire program that are assigned the value $null$ at a program point $l_1$ and dereferenced later at $l_2$ while ignoring the feasibility of the path from $l_1$ to $l_2$. For each $v\in V$, we then use the dereference on $v$ at $l_2$ as the criterion to compute a program slice~\cite{WeiserSlice} in which the null pointer dereference is guaranteed to be triggered under the same condition as it is in the original program. Finally, we regard all such slices to be the test code for null pointer dereference. As we explained above, the goal of this analysis is not to miss any potential bugs in the entire program. Therefore, the precision of the analysis is not a concern because the machine learning model, which is at the core of \met, is designed precisely for separating real bugs from false alarms.
Interested readers may refer to the supplemental material (Section~\ref{append:rules}) for the details of the data-flow analysis for some common types of bugs.

\vspace{3pt}
\noindent
\textbf{\circled{8}} \textbf{\textit{Prediction Interpreter}}
(Section~\ref{subsec:explain}).\,
After we identify the test programs for each type of bugs, \met first computes the raw embedding (\ie $g_\theta(\cdot)$) and then the relational embedding (\ie $\delta_{\theta}^{k}(\cdot)$) of each test program. Next, \met ranks test programs based on how distant their embeddings is to the mean point of this test set. Since in test-time the vast amount of the code \met sees will be bug-free, the mean point reflects the characteristic of the main cluster formed by the correct code, therefore, following the methodology of the training of \met, programs that are distant to the mean point are deemed to be the bugs inconsistent with the correct code.

Given test programs that \met predicts to be buggy, we proceed to explain its predictions. In general, we utilize the global attention mechanism as the basis of our explanations. In particular, we set out to find, in the buggy program, the path of statements that receive the highest global attention scores. Our intuition is that because of the goal of \met's global attention mechanism --- emphasizing the most essential part of a program graph that separates bugs from correct code, and a fact about the test code --- the inconsistent program constructs are the constant differentiator between buggy and correct programs, the statements that are highly attended will be parts of the program where the inconsistency lies. Technically, we compute a buggy trace out of those statements to explain the prediction that \met makes for the buggy programs. This is a standard practice we follow from the classic static analysis tools, which ensures that \met is as useful as static analysis tools regarding bug reporting. In Section~\ref{subsec:explain}, we explain in detail how to find the path covering the highest attended statements in the buggy program efficiently.
\section{\met's Graph Learning Approach}
\label{sec:gl}

In this section, we present \met's learning approach, in particular, we give the technical details for the key components in the training and test of \met.

\subsection{Definitions}

First, we define a standard bug detector. Conceptually, a bug detector is a model that assigns a label to an input program. 

\begin{definition}(\textit{Bug Detector})\label{def:detector}
	A bug detector is a K-way classifier $M:x \rightarrow \rho \in \mathbb{R}^{K}$ that for each label class predicts the probability $\rho_i\in[0,1]$ (\st $\sum_{i=1}^K \rho_i = 1$) that a program $x$ is a member of the class $i$.
\end{definition}

Definition~\ref{def:detector} does not exclude $P$ for being a correct program since bug free can also be a prediction class. Given such a bug detector, the label of $x$ is $L$ \st $L=\argmax_i \rho_i$ where $L \in [1,K]$, meaning, it can not predict any label outside of the $K$ classes provided in the training set. Now we define \met as follows:

\begin{definition}(\textit{Meta Bug Detector})\label{def:mbd}
	A meta bug detector is a ranking system $T$ for a set of programs $X$ \st $T:\forall x\in X(x,X/\{x\}) \rightarrow \upsilon\in \mathbb{R}$ that produces a scalar $\upsilon$ for $x$, which is the distance between $x$ and the mean point of $X$ in the embedding space.
\end{definition}

In essence, \met only examines how $x$ contradicts with or conforms to the other programs in $X$, thus
the label of $x$ is irrelevant. As a result, \met is capable of catching the type of bugs that are unobserved during training, a property that standard bug detectors do not possess. Interpreting \met's results is also straight-forward: programs that have a larger distance $\upsilon$ are more likely to be buggy, and the label of the bug is already determined from the data-flow analysis introduced in \textbf{\circled{7}} \textbf{\textit{Test Data Collector}} in Section~\ref{sec:over}.

\begin{definition}(\textit{Correctness of \met})\label{def:correct}
	Suppose all bug-free programs are assigned a label $\epsilon$. \met's prediction for a bug-free program $P$ is correct if \met predicts the label of $P$ to be $\epsilon$. Suppose a program $Q$ is buggy and for each bug $\mu$ in $Q$, there exists a sequence statements $T$ \st the execution of $T$ with an input $i$ triggers $\mu$ (denoted by $\mathit{exe}(T,i)=\mu$). Let $\bar{T}$ be the global minimum trace for $T$ \ie $\nexists \tilde{T} (\mathit{exe}(\tilde{T},i)=\mu) \land (\lvert \tilde{T} \rvert < \lvert \bar{T} \rvert)$ where $\lvert \cdot \rvert$ denotes the length of a sequence. \met's prediction for $Q$ is correct if \met outputs a sequence of statements $T'$ for each $\mu$ \st $\bar{T} \preccurlyeq T' \preccurlyeq T$ where $\preccurlyeq$ denotes subsequence and read as `is a subsequence of'.
\end{definition}

Our intuition is that $\bar{T}$ represents in theory the gold standard trace for explaining the bug $\mu$ because executing $\bar{T}$ with some input triggers $\mu$, and every statement in $\bar{T}$ is necessary. Therefore, if \met outputs a sequence $T'$ that subsumes $\bar{T}$, then $T'$ is a correct explanation for $\mu$. Since we design our analysis for collecting test programs to be sound, those test programs may have false positives (\eg statements that $\mu$ depends on are in fact unnecessary to the occurrence of $\mu$), therefore, $T'$, a path of one such test program, may also include false positives. This is the reason we do not require $T'$ to be identical to $\bar{T}$. Regarding $T'$ being a subsequence of $T$, this is because the analysis ruled out true negatives in $T$ (\eg statements completely unrelated to $\mu$) that no longer appear in $T'$. For interested readers, we left an example in the supplemental material (Section \ref{append:explanationexample}) to further illustrate the validity of Definition~\ref{def:correct}.



\subsection{An Interprocedural Program Dependence Graph}
\label{subsec:graph}

We design \met to be an interprocedural bug finder, in particular, we extend our PDG-based \textbf{\circled{2}} \textbf{\textit{Program Representation}} in Section~\ref{sec:over} to include procedures and procedure calls. We use ``procedure'' as a generic term referring to both the main function and the auxiliary procedures when the distinction between the two is irrelevant.

%
%

 \begin{figure}[t!]
 	\centering
 	\setlength{\fboxrule}{0.45pt}
 	\setlength{\fboxsep}{0pt}
		\begin{tabular}{|l|l|}
 		\hline
 		\scriptsize\ttfamily\bfseries{Auxiliary Procedure:}
 		&
 		\;\;\scriptsize\ttfamily\bfseries{Main Procedure:}
		\\
 		\begin{subfigure}{0.32\textwidth}
 			\lstset{style=mystyle, frame=0}
\begin{lstlisting}[morekeywords={Object, String},basicstyle=\scriptsize\ttfamily\bfseries]
void swap(int[] arr, int i) {
  int temp = arr[i];
  arr[i] = arr[i + 1];
  arr[i + 1] = temp;
  
}
\end{lstlisting}
 		\end{subfigure}
 		&
   	\begin{subfigure}{0.54\textwidth}
 			\lstset{style=mystyle, frame=0, xleftmargin=4pt}
\begin{lstlisting}[morekeywords={Object, String},basicstyle=\scriptsize\ttfamily\bfseries]
int[] bubbleSort() {
  int[] array = new int[]{3,5,1,8};
  for (int i = 0; i < array.length; i++)
    for (int j = 0; j < array.length - 1 - i; j++)
      if (array[j + 1] > array[j]) swap(array, j);
}
\end{lstlisting}
 		\end{subfigure}
 		\\\hline
 	\end{tabular}
 	\caption{A program with two procedures.}
 	\label{fig:intercode}
	
 \end{figure}
  
\begin{figure}[t!]
	\centering
	\includegraphics[height=7cm]{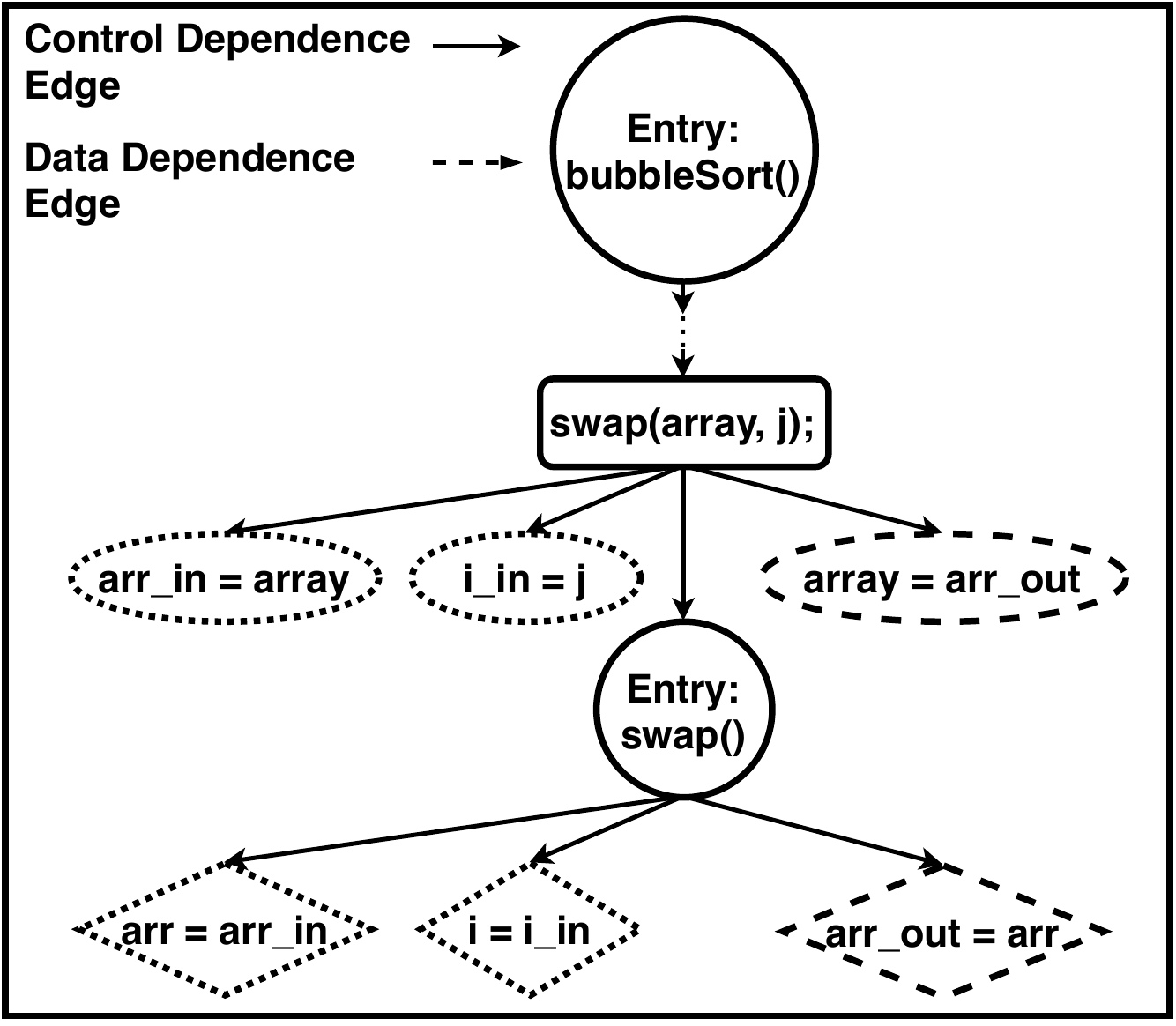}
	\caption{The new vertices --- call site (rectangle), actual in (dotted ovals), actual out (dashed ovals), formal in (dotted diamond), and formal out (dashed diamond) --- added to the interprocedural program dependency graph of the program in Figure~\ref{fig:intercode}, in particular, they represent parameter passing between two procedures \texttt{bubbleSort} and \texttt{swap}.}
	\label{fig:bubblesort}
	
\end{figure}

In simplest terms, an interprocedural  PDG includes a program dependence graph, which represents the main function, procedure dependence graphs, which represent the auxiliary procedures, and additional edges, which represent direct dependences between a call site and the called procedure. Regarding the design of the representation of parameter passing between procedures, we adopt the approach proposed in~\citet{10.1145Hor}. Specifically, we incorporate five new kinds of vertices to our PDG-based program graphs. A call site is represented using a call-site vertex; information transfer is represented using four kinds of parameter vertices. On the calling side, information transfer is represented by a set of vertices called actual-in and actual-out vertices. These vertices, which are control dependent on the call-site vertex, represent assignment statements that copy the values of the actual parameters to the call temporaries and from the return temporaries, respectively. Similarly, information transfer in the called procedure is represented by a set of vertices called formal-in and formal-out vertices. These vertices, which are control dependent on the procedure’s entry vertex, represent assignment statements that copy the values of the formal parameters from the call temporaries and to the return temporaries, respectively. Using the program in Figure~\ref{fig:intercode}, Figure~\ref{fig:bubblesort} depicts an example of the five new kinds of vertices. The rectangle represents the call site vertex.
Ovals with dotted/dashed lines represent the actual in/out vertices whereas diamonds with dotted/dashed lines represent the formal in/out vertices. Note that we do not consider the actual-out and formal-out vertices if the corresponding parameter is passed by value like the parameter \texttt{j} passed to the \texttt{swap} function in Figure~\ref{fig:intercode}. As before, all the new vertices are also represented by ASTs in the interprocedural PDG.

As for data dependences between procedures, they flow from actual-in vertices to formal-in vertices, and from formal-out vertices to actual-out vertices. Connecting procedure dependence graphs to form an interprocedural program dependence graph is straightforward, involving the addition of three kinds of edges: (1) a control dependence edge is added from each call-site vertex to the corresponding procedure entry vertex; (2) a data dependence edge is added from each actual-in vertex at a call site to the corresponding formal-in vertex in the called procedure; (3) another data dependence edge is added from each formal-out vertex in the called procedure to the corresponding actual-out vertex at the call site. For interesting readers, we left the entire interprocedural program dependence graph of the program in Figure~\ref{fig:intercode} to the supplemental material (Section~\ref{append:interprocedural}).

Another issue in constructing an interprocedural  program dependence graph is identifying, at a given method invocation, which method is being invoked amid dynamic dispatch or virtual function mechanism. This will not be a problem for programs used in the training phase because we know the exact paths along which bugs are triggered. For test code, we run interprocedural points-to analysis to track the origin of an object (\ie its allocation site) on which a virtual method is invoked. Technically, we perform flow and context-sensitive Andersen style points-to analysis with call-site sensitivity on the entire codebase. If the analysis returns multiple origins of an object, for each one, we create a procedure dependence  graph to represent its version of the method implementation, and then connect all procedure graphs to the program dependence graph of the calling method.

\subsection{Global Graph Attention}
\label{subsec:atten}

As a result of our adoption of the interprocedural  program dependence  graph, the size of program graphs increase, which makes GNN's message-passing mechanism, an already expensive procedure, even more costly.    
To overcome this challenge, we propose a global attention mechanism which helps to improve GNN's learning precision in meta bug detection. Our insight is that a few statements already sufficiently display the inconsistencies between correct and buggy code. Therefore, GNN should learn to emphasize such nodes
and downplay the rest in the graph. We realize this idea based on the classic attention mechanism~\cite{Bahdanau2015,luongetal15,NIPS20173f5ee243}. Below, we first present the technical details of the global graph attention, and then we explain how it makes GNN more effective in detecting the inconsistencies between correct 
and buggy code.

We first add to the graph another node, called \textit{meta node}, which connects every native node in the graph via a new type of edge. Next, we compute an attention score (Equation~\ref{equ:align1}) for each native node by taking the dot product between its embedding and the embedding of the \textit{meta node} $H$. Equation~\ref{equ:align2} then converts the attention scores into attention weights which can be deemed as a quota that limits the role each node plays in the message-passing procedure, in particular, Equation ~\ref{equ:gzmes1} presents how the message ${m}_n$ for node $n$ is produced after the attention weight of each of its neighboring nodes, ${\alpha}_u$, is taken into account; Equation ~\ref{equ:gzmes2} shows how node $n$'s representation is updated with its own attention weight ${\alpha}_n$.
\vspace*{-5pt}

\vspace*{-3pt}

\noindent\begin{minipage}{.5\linewidth}
	\begin{equation}
		e_n =       \,h_{n} \odot H   \,\,\,\,\,\,\,\,\,                     \label{equ:align1}
	\end{equation}
	\end{minipage}%
\begin{minipage}{.5\linewidth}
	\vspace{8pt}
	\begin{equation}
		\alpha_n =  \frac{\mathit{exp}(e_n)}{\sum_{i=1}^{|N|}\mathit{exp}(e_i)}  \,\,\,\,\,\,  \label{equ:align2}
	\end{equation}
	\end{minipage}

\noindent\begin{minipage}{.5\linewidth}
	\vspace{5pt}
	\begin{equation}
	 {m}_n        = \sum_{\mathclap{u\in \Omega_n}} \alpha_u f(h_{u}) \label{equ:gzmes1} 
	\end{equation}
	\end{minipage}%
	\begin{minipage}{.5\linewidth}
	\vspace{-5pt}
	\begin{equation}
		h'_{n}      = \mathit{GRU}({m}_n, \alpha_n h_{n})             \label{equ:gzmes2}
	\end{equation}
	\end{minipage}
	
\vspace*{4pt}
Initially, the embedding of the \textit{meta node} is assigned randomly. After a round of message-passing, we update it as follows:
\begin{equation}
H' = \sum_{n\in V}\alpha_n h_n \nonumber
\end{equation}
where $V$ denotes the set of native nodes in the graph.



To achieve \met' learning objective, or in essence minimize GNN's training loss, GNN will inevitably attend to the nodes that display the inconsistency between buggy and correct program graphs since they are the biggest differentiators of the two sets of graphs. In other words, even if some buggy programs substantially differ from some correct programs in surface-level textual/syntactic features, given nodes that display the inconsistencies between the two sets of graphs are the constant difference maker for every buggy and correct program, GNN will nudge toward them in order to minimize its training loss. While such nodes are receiving higher attention weights, the \textit{meta node} starts to reflect their characteristics, as a result, more attention will be given to those nodes in later rounds of message-passing, subsequently, GNN separates the two sets of graphs even further, and the virtuous circle carries on. 

We note that our global graph attention differs from the widely used graph attention mechanism (\eg GAT~\cite{velikovi2018graph}). The goal of global attention mechanisms is to emphasize nodes that reflect the essential characteristics of the entire graph while downplaying the rest. We realize this goal by assigning an attention weight for every node based on how it relates to every other node in the graph (\eg through \textit{meta node}). In contrast, GAT prioritizes nodes only within the neighborhood of a node (denoted by node $n$). The attention weight of each neighboring node (denoted by node $u$) is purely determined by node $n$ and $u$. It is rather unclear how such a localized attention mechanism can be applied to prioritize nodes within the entire graph. Although graph pooling~\cite{cangea2018towards,9432709,lee2019self} share some similar characteristics with the global attention mechanism, they are designed for a different purpose: preventing models from overfitting by reducing the size of data representations. For this reason, they cut nodes from graphs in an explicit manner; and thus require users to specify the number of nodes to be retained in the graph via a hyperparameter. In contrast, our global graph attention aims to improve the precision of GNN, and it does so by highlighting the key nodes without altering the graph structure. 



\subsection{\met's Loss Function}
\label{subsec:loss}



First, we give an overview of the triple loss function from which we derive \met's loss function. Consider a setting where the training data is prepared into a set of triples ($\bar{x}$, $x_{+}$, $x_{-}$) where $\bar{x}$ is an anchor input, $x_{+}$ is a positive input of the same class as $\bar{x}$, $x_{-}$ is a negative input of a different class from $\bar{x}$. The objective of triplet loss models is to embed each triple in the way that samples with the same label, $\bar{x}$ and $x_{+}$ are smaller in distance than those with different labels, $\bar{x}$ and $x_{-}$. {The equation} below formulates this concept. 
\begin{equation}
\label{equ:loss}
{L}(\bar{x},x_{+},x_{-}) = \, \max(\lVert \phi(\bar{x})-\phi(x_+) \rVert^2 - 
                        \lVert \phi(\bar{x})-\phi(x_-) \rVert^2 + \epsilon, 0)    \nonumber
\end{equation}
where $\lVert \cdot \rVert^2$ denotes the Euclidean distance function and $\phi(\cdot)$ is the embedding function. 

Clearly, ${L}(\bar{x},x_{+},x_{-})$ equals to $0$ when the distance between $\bar{x}$ and $x_{-}$ is greater than that between any $\bar{x}$ and $x_{+}$ by at least $\epsilon$. The purpose of $\epsilon$ is that, when the representations produced for $\bar{x}$ and $x_{-}$ are distant enough, further training will save the effort from continuing to enlarge that distance, instead, it can focus on other more difficult triplets. Following the convention in the literature, we set the value of $\epsilon$ to be 1.

Summing over all triplets in the training set, we obtain the triplet loss function:
\begin{equation}
\label{equ:loss1}
\mathcal{L} = \sum_{i=1}^{N} L(\bar{x}^i,x_{+}^i,x_{-}^i) \nonumber
\end{equation}

Now, we can instantiate the above equation to create \met's loss function according to its learning objective (Equation~\ref{equ:trainsema}). To sort our training programs into a set of triplets, we first create a set of pairs by pairing each buggy program with each correct program for each type of inconsistency. Then we use the average embedding of those programs as the anchor for each pair. By substituting $p$, a correct program, for $x_{+}$, $q$, a buggy program, for $x_{-}$ and $\bar{r}$ for $\bar{x}$, we create \met's loss function for one type of inconsistency below:
\begin{equation}
\label{equ:MBDloss}
\mathcal{L}(P,Q) = \sum_{q \in Q}\sum_{p \in P} \max(\lVert \bar{r}-\delta^{k}_\theta(q) \rVert^2 - \lVert \bar{r}-\delta^{k}_\theta(p) \rVert^2 + \epsilon, 0) \nonumber
\end{equation}
where $\bar{r}$ is defined in Equation~\ref{equ:referenceset} and $\delta^{k}_\theta(\cdot)$ is defined in Equation~\ref{equ:rela}. Summing losses over all types of inconsistencies, we obtain \met's loss function:
\begin{equation}
\label{equ:MBDlossRA}
\mathcal{L}(\mathcal{U}) = \sum_{U \in \mathcal{U}} \mathcal{L} (U_-,U_+)
\end{equation}
where $\mathcal{U}$ denotes the entire set of training programs, $U$ is the training programs for one type of inconsistency, $U_-$/$U_+$ represents the buggy/correct programs in $U$. 

\subsection{Explaining \met's Predictions}
\label{subsec:explain}

In this section, we give a detailed presentation on \met's self-explainability, specifically, how \met produces a buggy traces out of the buggy program. We omit \met's procedure for detecting buggy programs during test, which is already presented in \textbf{\circled{8}} \textbf{\textit{Prediction Interpreter}} in Section~\ref{sec:over}. 

%


As we explained earlier, the global attention mechanism in principle would attend to program statements that constitute the inconsistency between buggy and correct code. However, directly presenting the highest attended statements in the buggy program as explanations lacks clarity, and thus, hinders the utility of \met. Instead, we set out to find the program path in the buggy program that contains the highest attended statements, the same way static analysis tools educate developers when they report bugs. 

First, we formulate the problem of trace generation more precisely: find a feasible program path along which statements add up to the highest attention scores. Note that a path satisfying this requirement is not guaranteed to trigger the bug. However, this should not refute our formulation of the problem. Because ultimately it's the model's fault for not learning the right features to describe the bug. Considering the interpreter's job is to merely present what the model has learned, we believe our formulation is reasonable. 

Given this problem formulation, an optimal solution requires an exponential time algorithm, instead, we propose a 1-basic block off algorithm that can solve this problem much more efficiently. The high-level key steps are as follows: (1) we automatically take into account statements that are integral to the type of the bug (\eg null assignment and dereferencing statement for null pointer dereference), in addition, we mark the criterion we use for slicing (described in \textbf{\circled{7}} \textbf{\textit{Test Data Collector}} in Section~\ref{sec:over}) to be the point at which the bug occurs. 
(2) we then consider 10 other statements that receive the highest attention scores. We select 10 because they together make up over 99 percent of the attention scores for any program in our extensive experiments; (3) out of the statements identified in step (1) and (2), we identify the path of the highest attention scores without considering the path feasibility, meaning, we ignore statements in an \texttt{if}/\texttt{else} clause only if the opposite \texttt{else}/\texttt{if} clause contains statements combined into a higher attention score. An exception, which in fact applies to the whole procedure, is that we never remove integral statements even if they have the lowest attention scores.
(4) Next, we check the feasibility of the path found in the previous step, if it is infeasible, we identify a new (\ie unexplored) path by deselecting statements with the lowest possible attention scores. In other words, the new path either circumvents the \texttt{if} branch of deselected statements (if it is a stand-alone \texttt{if} block) or takes the opposite branch to deselected statements; (5) if the new path is again unfeasible, we first recover the path found at step (3), and then deselect statements with the next lowest possible attention scores. This is the reason we name our algorithm 1-basic block off as the new path identified always covers exact 1 different basic block from the original path identified in step (3); (6) we repeat step (5) until we find a feasible program path or deselected all statements identified in step (2). In theory, this algorithm may not find the path that satisfies the formulation above, however, this is not a  concern because the formulation should not be deemed as a requirement --- because solutions guaranteed to satisfy the formulation may still violate Definition~\ref{def:correct} --- 
but rather a guidance that leads \met to discover the intended path. In practice, the algorithm works well and the reason is simple: the global attention weights are already accurate reflection of the buggy trace, if anything, it only needs minor corrections which is precisely what this 1-basic block off approach does.

\begin{algorithm}[!tb]

	\algnewcommand\Sz{size\xspace}
	\algnewcommand\Md{method\xspace}
	\algnewcommand\Mn{method.name\xspace}		
	\algnewcommand\Mb{method.body\xspace}	
	\algnewcommand\Mp{method.parameters\xspace}	
	\algnewcommand\Mh{method.head\xspace}	
	\algnewcommand\Mr{method\_header\xspace}		
	\algnewcommand\Sd{new\xspace}	
	\algnewcommand\Sr{new Method(remaining\_statements, method.head)\xspace}		
	\algnewcommand\Ml{model\xspace}
	\algnewcommand\Lb{label\xspace}
	\algnewcommand\Pd{result\xspace}	
	\algnewcommand\K{k\xspace}
	\algnewcommand\I{i\xspace}
	\algnewcommand\S{s\xspace}
	\algnewcommand\St{selected\_statements\xspace}		
	\algnewcommand\Rs{remaining\_statements\xspace}
	\algnewcommand\Ts{statements\xspace}
	\algnewcommand\Ti{statements[i]\xspace}	
	\algnewcommand\Tz{statements.size\xspace}
	\algnewcommand\To{statements.size-1\xspace}		
	\algnewcommand\Tr{traversed\_statements\xspace}
	\algnewcommand\Se{fragment\xspace}	
	\algnewcommand\Rm{region[-1]\xspace}	
	\algnewcommand\Mm{m\xspace}
	\algnewcommand\Ic{is\_continue\xspace}
	\algnewcommand\Nm{m\xspace}
	\algnewcommand\Ci{index\xspace}
	\algnewcommand\Cd{depth\xspace}
	\algnewcommand\Ip{index++\xspace}
	\algnewcommand\Dp{depth++\xspace}
	\algnewcommand\Me{max\_depth\xspace}
	\algnewcommand\Sa{selected\xspace}			
	\algnewcommand\Lz{selected.size\xspace}
	\algnewcommand\rec{Reconstruct\xspace}
	\algnewcommand\vef{VerifyCodeFragment\xspace}
	\algblockdefx[Foreach]{Foreach}{EndForeach}[1]{\textbf{foreach} #1 \textbf{do}}{\textbf{end foreach}}

	\caption{Find a feasible path with the highest attention score.}
	\label{alg:explainpredictions}
  \begin{algorithmic}[1]
  \Procedure{FindFeasiblePath}{program, attention\_score}
		\LineComment{\ \ \ \ \(\triangleright\) Get the integral statements from the program.}
		\State integral\_statements $\gets$ \textit{GetIntegralStatements}\!\:(program)\label{line:integral}
		\LineComment{\ \ \ \ \(\triangleright\) Get top\hspace{.3mm}-10 highest-attended statements.}
    \State statements $\gets$ \textit{ReturnStmtsWithTopNScore}\!\:(program, attention\_score, 10)
		\State statements.\textit{Incorporate}\!\:(integral\_statements)
		\LineComment{\ \ \ \ \(\triangleright\) Calculate the scores of the branches covering these statements.}
		\State branches_score_map  $\gets$ \textit{CalcBranchScores}\!\:(statements, attention\_score)
		\LineComment{\ \ \ \ \(\triangleright\) Select the path with the highest score and get its branches.}
		\State path, branches $\gets$ \textit{SelectPathAndBranches}\!\:(branches_score_map)
		\LineComment{\ \ \ \ \(\triangleright\) Each candidate path is constructed from variable \textit{path} instead of the last candidate path.}\label{line:pathbranch}
		\State cand\_path $\gets$ path 
    \While{!\,\textit{IsPathFeasible}\!\:(cand\_path) \textbf{and} !\,branches.\textit{Empty}\!\:()}
			\LineComment{\ \ \ \ \ \ \ \ \(\triangleright\) Pop the lowest score branch in the branch set.}
			\State lowest\_score\_branch $\gets$ \textit{PopBranchWithLowestScore}\!\:(branches)
			\State opposite_branches $\gets$ \textit{FindOppositeBranches}\!\:(lowest_score_branch) 
			\If {opposite_branches.\textit{Empty}\!\:()} \Comment{This represents a single \texttt{if} block.}
			\State cand\_path $\gets$ \textit{RemoveBranch}\!\:(path, lowest\_score\_branch) \Comment{Remove the \texttt{if} block.}
			\Else \Comment{This represents a \texttt{if} statement with multiple branches.}
			\State cand\_path $\gets$ 	 \Call{FindSubpath}{path, lowest\_score\_branch, opposite_branches} \label{line:mutate} 
			\EndIf
    \EndWhile
    \State  \Return{cand\_path}
  \EndProcedure
  
  \item[]
		
	\Function{FindSubpath}{$path, branch, branches$} 
	\LineComment{\ \ \ \ \(\triangleright\) We enumerate all subpaths within the branches including their sub-branches.}
	\State subpaths $\gets$ \textit{EnumerateAllSubpathsInBranches}\!\:(branches)
	\While{!subpaths.\textit{Empty}\!\:()}
	  \State subpath $\gets$ \textit{PopSubpathWithHighestScore}\!\:(subpaths)
	  \LineComment{\ \ \ \ \ \ \ \ \(\triangleright\) Replace \textit{branch}'s subpath by \textit{subpath}.}
	  \State cand\_path $\gets$ \textit{ReplaceSubpath}\!\:(path, branch, subpath)  
      \If {\textit{IsPathFeasible}\!\:(cand\_path)} 
    \State  \Return{cand\_path}
	\EndIf
    \EndWhile

	\EndFunction
	
\end{algorithmic}
\end{algorithm}

Algorithm~\ref{alg:explainpredictions} gives the details. We first obtain the integral statements
according to the type of the bug. Next, we collect ten statements with the highest attention score via function \textit{ReturnStmtsWithTopNScore}. If some of those statements happen to reside in branches, we compute the score for a branch as the sum of its statements (\textit{CalcBranchAndScore}). Note that we compute the branch score in a recursive manner to properly handle the nested branches.
\textit{SelectPathAndBranches} then selects a path that covers the branches with the highest scores. If neither of the opposite branches contain any top\hspace{.3mm}-10 highest-attended statement, we randomly select a branch to cover. 
Then we determine whether or not the path is feasible, if not, we retrieve the branch in the path with the lowest attention score, and proceed with the following two scenarios. 
(1) if the retrieved branch is
a single \texttt{if} block, we just remove it; (2) if there are other branches associated with the retrieved branch (\eg one or multiple \texttt{else} blocks), we replace the retrieved branch with a subpath within an associated branch which makes the entire path feasible. In particular, we enumerate all subpaths within each associated branch (taking into account the nested branches) through \textit{EnumerateAllSubpathsInBranches}, and look for the subpath with the highest score (\textit{PopSubpathWithHighestScore}) to replace the retrieved branch (\textit{ReplaceSubPath}). The first one we find that produces a feasible path is declared as the buggy trace. 
The algorithm will always terminate after, in the worst case, it has dealt with all branches in the path discovered at line~\ref{line:pathbranch}. 

As a common practice, we perform under-constrained symbolic execution~\cite{Engler2831147} to determine the feasibility of a path. Specifically, we start executing a test program from the entry point of the procedure that contains the first statement (\eg an integral statement or a top-10 attended statement) and unroll every loop once to reduce the number of paths to be explored.

Concerning the usability of \met, we decorate the buggy trace as follows. First, we mark the statement at which the bug occurs within a bounding box, and cut off all subsequent statements in the trace. Second, we number statements in the trace according to the order in which they are executed. Finally, we underline a few highest attended statements as key steps of the trace.

\section{Evaluation}
\label{sec:eva}

We implement \met into \tool, and evaluate it on three benchmarks including the well-established datasets like Defects4J ~\cite{10.1145/2610384.2628055} and JBench ~\cite{10.1145/3196398.3196451}. We also examine in-depth the results of \met's to reveal the strengths and weaknesses of \met. Finally, we set out to confirm the necessity of each key component of \tool. 

\subsection{Implementation}
Due to the ample availability of Java datasets, we target a variety of bugs in Java programs. For a different language, \tool only requires a front-end analyzer that extracts program dependence graphs for programs in the new language, the rest of the workflow is language-agnostic. We use DECKARD~\cite{jiang2007deckard} to detect syntactic similar programs. We construct program dependence graphs using Spoon~\cite{pawlak:hal-01169705}, an open-source library for analyzing Java source code. We use Doop~\cite{Doop} for pointer analysis. We realize the prediction interpreter on top of angr~\cite{7546500}, a fairly mature framework that supports the symbolic execution in Java. We realize \tool in Tensorflow based on GGNN’s implementation. All experiments are performed on a 3.7GHz i7-8700K machine with 32GB RAM and NVIDIA GTX 1080 GPU.

\begin{wraptable}{R}{0.5\textwidth}
	\vspace{-14pt}
	\captionsetup{skip=1pt}
	\caption{Details about our dataset.}	
	\centering
	\resizebox{.5\textwidth}{!}{
	\begin{tabular}{r|cc}
		\hline
		\multirow{2}{*}{Bug Types} &   \multicolumn{2}{c}{\# Programs}  \\
		\cline{2-3}
		&\tabincell{c}{Buggy} & \tabincell{c}{Correct}  \\\hline
		Null pointer dereference (NPD) & 35 & 287   \\\hline
		Number format exception (NFE) & 16 & 149   \\\hline
		Array index out-of-bound (AIO) & 23 & 216  \\\hline
    	Stack overflow exception (SOE) & 14 & 140   \\\hline
		File handle leak (FHL)& 11 & 110           \\\hline
		\multicolumn{1}{r|}{Total}& 99  & 992 \\\hline
	\end{tabular}
	}
	\label{tab:stas}
	 \vspace{-3pt}
\end{wraptable}



\subsection{Proof-of-Concept}
First and foremost, we conduct a proof-of-concept for meta bug detection. That is can \tool at the very least separate bugs, the kind of which it has never observed during training, from the correct code that we automatically mined for our dataset.


\vspace*{3pt}
\noindent
\textbf{\textit{Dataset.}}\, We build an automated pipeline (presented in \textbf{\circled{1}} \textbf{\textit{Training Data Collector}} in Section~\ref{sec:over}) to curate the dateset. We collect bugs from Bugs.jar~\cite{saha2018bugs} and BugSwarm~\cite{Tomassi2019}, two mature datasets of real bugs in Java programs. A caveat is that both datasets trace bugs from exceptions that programs throw at the run time, which can induce noise to their data. As an evidence, we notice some exceptions are thrown independently of the program that the datasets include as buggy. For example, some bugs in Bugs.jar are linked to the \textit{FileNotFoundException}, which is thrown when the file to be accessed is not found in the system. After a careful investigation on the root cause of those bugs, we conclude that they are unrelated to programming mistakes. In some cases, it's due to the lack of sufficient privilege to write the file at the first place. A few others are caused by the network issues that prevent the file to be transmitted from a remote server. Also, there are exceptions triggered due to the violation of the project-specific assumptions. For example, in reality APIs are never called in the order specified in the test code; or input arrays are always of a certain size. Furthermore, we found some exceptions are triggered by the runtime environment which do not have static features (\eg \textit{OutOfMemoryException}). Finally, we only consider bugs that are reproducible through the automated scripts provided by the dataset. We 
%
sort bugs in our dataset into five categories: 
null pointer dereference, number format exception, array index out-of-bound, stack overflow exception, and file handle leak --- under which bugs are guaranteed to be valid. 
For reader's convenience, we describe number format exception and stack overflow exception since others are well-known. A stack overflow exception is thrown when the amount of call stack memory allocated by the JVM is exceeded. This is normally caused by the excessive deep or infinite recursion. A number format exception occurs when a string (with improper format) can not be parsed into a number. 
We pair each bug with ten correct programs such that in most cases every correct program still maintains the inconsistency with the bug. The reason that correct programs are not exact 10 times as many as bugs is that bugs extracted from the same project may share common correct programs.
Table \ref{tab:stas} gives the statistics of our dataset.

\vspace{3pt}
\noindent
\textbf{\textit{Cross Validation.}}\, To evaluate the effectiveness of \tool, we conduct a cross-validation on our dataset. We withdraw programs for each type of bugs from the entire dataset in turn as the validation set, and leave the rest as the training set. Our goal is to test if \tool can indeed catch the type of bugs it has never seen during training. Concerning the validity of the experiment, we manually confirm that the type of inconsistency overlaps between the validation set and the training set for every type of bugs.

\vspace{3pt}
\noindent
\textbf{\textit{Metric.}}\, We emphasize again that we examine \met's performance strictly according to Definition~\ref{def:correct} throughout our evaluation. Because \tool is, in essence, a ranking system of test programs, we define a cut-off point $N$ as the separation between buggy and correct programs. In other words, we consider the top $N$ programs to be buggy and inspect how many bugs are covered (denoted by TP), how many bugs are missed (denoted by FN), and how many false alarms are reported (denoted by FP). The value of $N$ can be set according to developers' intention whether they favor \met to produce less false warnings or identify more bugs. For instance, $N=1$ can represent an opportunistic approach developers tend to adopt due to their unwillingness for an extensive bug inspection. $N=5$ (or even a larger number), on the other hand, represents a conservative approach in which all bugs should have been caught because developers will not inspect reports further down the list.
Considering this evaluation only serves as a proof-as-concept, we simply set $N$ to be the number of buggy programs \wrt the type of bugs withdrawn to the validation set. 

\vspace{3pt}
\noindent
\textbf{\textit{Baselines.}}\, We use GGNN, the base model of \tool, and GINN~\cite{Wang3428205}, one of the most recent advances in learning-based bug detection, as the baselines for this experiment. Because both \tool and GINN are built upon GGNN, the three models have the same set of hyperparameters. For a fair comparison, we use the default values provided by GGNN's implementation for all models. We train GGNN and GINN to predict only two labels regardless of the type of the bugs: buggy or correct. To counter the lopsided ratio between the correct and bug code in the training set, we give six times higher weights to the loss that models incur on buggy programs than correct programs so that models will prioritize the minimization of its loss on buggy code over correct code. Since GGNN and GINN directly learn bug patterns, we expect them to perform poorly on this dataset.

\begin{table*}[t]
	\captionsetup{skip=1pt}
	\caption{Results of cross validation. The acronyms of bug types are introduced in Table~\ref{tab:stas}.}
	
	\begin{tabular}{l|c c c|c c c|c c c|c c c|c c c}
		\hline
		\multirow{2}{*}{}
		& \multicolumn{3}{c|}{NPD} & \multicolumn{3}{c|}{NFE} & \multicolumn{3}{c|}{AIO} & \multicolumn{3}{c|}{SOE} & \multicolumn{3}{c}{FHL} \\\cline{2-16}
		& TP & FP & FN & TP & FP & FN & TP & FP & FN & TP & FP & FN & TP & FP & FN \\\hline
		GINN  & 8& 36& 27& 2& 12& 14& 2& 15& 21& 3& 18& 11& 2& 15&9 \\\hline
		GGNN  & 4& 42& 31& 2& 14& 14& 1& 18& 22& 3& 16& 11& 1& 18&10 \\\hline		
		{\tool} & {24}& {11}& {11}& 
		{11}& {5}& {5}& 
		{14}& {9}& {9}& 
		{9}& {5}&{5}& 
		{7}& {4}& {4} \\\hline
	\end{tabular}
	\label{tab:crossvalidation}
	\vspace*{-5pt}
\end{table*}

\vspace{3pt}
\noindent
\textbf{\textit{Results.}}\, Table~\ref{tab:crossvalidation} shows \tool's results in cross validation. Clearly, \tool is quite precise across all types of bugs, yielding around 0.35 false positive/negative ratio, which is significantly better than GGNN and GINN. This study confirms that \tool, which learns the manner in which buggy code is inconsistent with correct code, can detect the type of bugs that are unobserved during training. On the contrary, existing models, which learn specific bug patterns, can not deal with the type of bugs which they are not trained on.

\subsection{Evaluating \tool on Defects4J}
\label{subsec:def}

In this experiment, we evaluate the effectiveness of \tool in catching bugs from real-world codebases. This is the typical way static bug finders are used except in our case bugs are already identified. For groundtruth, we extract 79 bugs from Defects4J~\cite{10.1145/2610384.2628055}, arguably the most well-established dataset of bugs in Java programs. Our bugs come in three different types: null pointer dereference (39), array index out-of-bound (34), and number format exception (6). 
In addition to reasons for which we can not use all bugs in Bug.jar, we also exclude bugs in Defects4J that are related to the test suite themselves (\eg test code written in JUnit). We ignore them because they are far less significant than bugs occurred in the real codebases. For null pointer dereference, the 39 bugs used in our evaluation is a superset of the 26 bugs that~\citet{Tomassi2021} extract from Defect4J for their evaluation.

\vspace{3pt}
\noindent
\textbf{\textit{Baselines.}}\, Although Defects4J has been the standard benchmark for evaluating fault localization works~\cite{Zou2021,MASRI2015103}, we do not choose them as baselines because they are predominately powered by dynamic approaches that assume the pre-existence of test cases that can trigger the bug. It is clear such test cases are not available in the setting of static bug finding. Instead, we pick baseline methods based on an extensive study conducted by~\citet{Tomassi2021} on the real-world effectiveness of several noteworthy static bug finders, especially that they also use Defects4J as an evaluation benchmark. According to their results, we
select the top performers: \eradicate, which find the most bugs, and Infer, which has the highest precision.
\eradicate is a type checker that checks for @Nullable annotations in Java programs by performing a flow-sensitive analysis to propagate null-related information through assignments and calls. 
Infer, a prominent static-analysis tool developed by Facebook, uses bi-abduction analysis to find bugs. Despite being built upon sound programming language theory~\cite{10.1145/2049697.2049700}, Infer is an unsound tool in practice.~\citet{Tomassi2021} discuss in detail sources of Infer's unsoundness (at Table II). Nonetheless, our experiments set aside the soundness notion while focusing on the effectiveness of each tool in finding bugs in real-world programs. For a learning-based method, we pick FICS~\cite{263838}, arguably the state-of-the-art anomaly detection tool, as another baseline of this experiment. FICS is aimed at C/C++ which works on LLVM IR~\cite{LLVMIR}. To make it compatible with Java programs, we use JLang ~\cite{JLang} as a front-end to generate LLVM IR of Java programs, which then get passed to FICS. To ensure the correctness of our implementations, we perform differential testing on the behavior of two executable files: one is x86 machine code obtained via Java source code $\rightarrow$ LLVM IR $\rightarrow$ x86 machine code and the other is Java class file directly compiled from Java source code. Our implementation passes all test cases in the test suite provided by JLang. A key hyperparameter of FICS's is the code granularity at which the analysis is performed. We use 1-Con (\ie one basic block) and full-Con (\ie the whole data dependence graph) with a similarity threshold of 95\%, which gives the best performance of FICS on this dataset. We have also tuned the other hyperparameters to make sure FICS performs to the best of its ability. We compare \tool against \eradicate and Infer in catching null pointer dereference bugs, the only kind of bugs used in this experiment that \eradicate and Infer can handle. Since we did not find any static analysis tool in the literature that can catch array index out-of-bound or number format exception in a fully automated fashion, we only compare \tool with FICS on those two kinds of bugs.

\vspace{3pt}
\noindent
\textbf{\textit{Test Procedure.}}\, We follow \met's testing workflow depicted in Figure~\ref{fig:testing}.
%
At the end of the workflow, \tool produces a ranking of test programs \wrt each type of bugs. Given that \tool is evaluated in a real setting of static bug finding, we set $N$ to be 1 and 5 to mimic two contrasted approaches developers can adopt to inspecting bug reports. As we explained in the previous section, $N=1$ enables developers to focus on bugs that are highly likely to be buggy, alternatively, $N=5$ would let developers inspect all code that are potentially buggy.
The model used in this experiment is retrained with the entire dataset (Table~\ref{tab:stas}) with the same set of hyperparameters as before. 


\begin{table}[t]
	\captionsetup{skip=1pt}
	\caption{Comparing \tool against \eradicate, Infer and FICS on Defects4J. The acronyms of bug types are introduced in Table~\ref{tab:stas}. {\large\textbf{+}} is the number of bugs a tool finds but not others. When counting {\large\textbf{+}} for a baseline, we consider \tool in top-5 prediction mode which is \tool's the best performance. Because there does not exist a static analysis tool in the literature that can check AIOE and NFE in a fully automated manner, we only compare \tool against FICS. For interested readers, we have left a comprehensive, 1-on-1 bug comparison among all tools to the supplemental material (Section~\ref{append:onebyone}). We emphasize again all tools are unsound, however, this comparison sets aside the soundness issue while focusing on the effectiveness of each tool in catching bugs in Defects4J, the same way~\citet{Tomassi2021} evaluate static bug finders.}
	\begin{tabular}{c|c|c c c c|c c c c|c c c c}
		\hline
		\multicolumn{2}{l|}{}  & \multicolumn{4}{c|}{NPD} & \multicolumn{4}{c|}{AIO} & \multicolumn{4}{c}{NFE} \\\cline{3-14}
		\multicolumn{2}{l|}{}  & TP & FP & FN & \large{\textbf{+}} & TP & FP & FN & \large{\textbf{+}} & TP & FP & FN & \large{\textbf{+}} \\\hline
		 \multicolumn{2}{c|}{Infer} & 3  & 459  & 36 & 0 & - & - & - & - & - & - & - & - \\\hline
		 \multicolumn{2}{c|}{\eradicate} & 11 & 16476 & 28 & 6 & - & - & - & - & - & - & - & - \\\hline
		 \multicolumn{2}{c|}{FICS}  &  3  & 4182 & 36 & 2 & 6 & 3555 & 28 & 6 & 1 & 732 & 5 & 0 \\\hline
		\multirow{2}{*}{\tool} &  top-1 & 6  & 4  & 33  & 3 & 6 & 3 & 28 & 6 & 2 & 1 & 4 & 1 \\\cline{2-14}
		& top-5 &{11}  & {39}  & {28} & {8} & {10}  & {35}  & {24} & {10} & {4}  & {11}   & {2} & {3} \\\hline
	\end{tabular}
	\label{tab:compareres}
\end{table}

\vspace{3pt}
\noindent
\textbf{\textit{Results.}}\, Considering \eradicate and Infer are regarded as the state-of-the-art in static bug finding, we make a tremendous effort to confirm the correctness of their results in our evaluation.
First, we compare the results of \eradicate and Infer to those reported in~\citet{Tomassi2021} on the 26 bugs that both evaluations use. After a thorough inspection, we find the two sets of results indeed match. We describe the details of our inspection in the supplemental material (Section~\ref{append:manual}). Second, since \eradicate and Infer analyze the remaining 13 null pointer dereference in the exact same manner as they analyze the 26 shared bugs (\eg the same version of \eradicate and Infer, the same commands for invoking \eradicate and Infer), we believe their results are also correct. Finally, in case Infer have found new bugs not included in Defects4J, we manually examine Infer's 459 false reports. We rule out 428 on our own as clear false alarms, and left the developers the rest which are all declared to be false in the end. We did not manually inspect \eradicate's (or FICS's) false reports for two reasons: first, given the number of false alarms they produce, such a manual inspection will be extremely time-consuming; and second, even if Defect4J missed bugs for some projects, all tools will be affected the same way, thus, the results will not bias toward \tool. Below, we discuss the results of our evaluation.

For null pointer dereference, in top-1 prediction mode, \tool already finds more bugs than all baselines except \eradicate. In particular, it produces orders of magnitude less false alarms than all baselines. In top-5 prediction mode, \tool finds the same number of bugs as \eradicate but with 400+ times less false warnings. Results for array index out-of-bound and number format exception tell the same story: \tool not only finds more bugs than FICS, but has a substantially higher precision. Despite \tool's far superior effectiveness, we find that the baselines are in fact complementary to \tool. Specifically, \eradicate and \tool find only two common null pointer dereference while FICS does not overlap with \tool at all in array index out-of-bound bugs (more details can be found in Section~\ref{append:onebyone} in supplemental material). This finding suggest that each tool has its own strengths and weaknesses, and no tool is completely replaceable.


We use a buggy (Figure~\ref{fig:defect4jbug11}) and its reference program (Figure~\ref{fig:defect4jfix11}) to showcase how \tool detects null pointer dereference. The reference program is the correct program that receives the highest attention weight (Equation~\ref{equ:aw}) \wrt the embedding of the buggy program. As we explained in \textbf{\circled{6}} \textbf{\textit{Relational Embedding}} in Section~\ref{sec:over}, the program of the highest attention weight is the most indicative of the incorrectness of the bug, thus, we use it as a reference to illustrate the bug. The statements receiving the highest attention scores in both programs are highlighted in shadowbox. The inconsistency between the two programs is that the correct program always checks the object to be dereferenced (\eg line 9, 15, and 20) whereas the buggy program misses the check for the last dereference at line 26.

\begin{figure*}[tb]
	\centering
	\captionsetup{skip=4pt}

	\begin{subfigure}{0.37\textwidth}
		\lstset{style=mystyle, numbers = left, framexleftmargin=8pt}		
\begin{lstlisting}[morekeywords={Object, String}]
// From Chart-4/source/org/jfree/chart/
// plot/XYPlot.java:4493
public Range getDataRange(ValueAxis 
    axis) {
  ...
  if (isDomainAxis) {
    if (@r != null@) {
      result = Range.combine(result, 
          @r.findDomainBounds(d)@);
    }
    ...
  }
  ...

  else {
    if (@r != null@) {
      result = Range.combine(result, 
          @r.findRangeBounds(d)@);
    }
    ...
  }
  ...
  
  // an inconsistent null check, 
  // variable r may also be null here.
  @Collection c = r.getAnnotations();@
  ...
}
\end{lstlisting}
		\caption{A null pointer dereference.}
		\label{fig:defect4jbug11}
	\end{subfigure}
	\;\;\;\;
   	\begin{subfigure}{0.48\textwidth}
		\lstset{style=mystyle, numbers = left, framexleftmargin=8pt}		
\begin{lstlisting}[morekeywords={Object, String}]
// From Chart-5/distribution/source/org/jfree/chart/
// axis/CyclicNumberAxis.java:425
protected List refreshTicksHorizontal(Graphics2D g2,
    Rectangle2D dataArea,
    RectangleEdge edge) {
  CycleBoundTick lastTick = null;
  while (currentTickValue <= upperValue) {
    if (isVerticalTickLabels()) {
      if ((@lastTick != null@) && ...) {
        result.add(new CycleBoundTick(
          ..., @lastTick.getNumber@(), ...));
      }
    } else {
      if (edge == RectangleEdge.TOP) {
        if (@lastTick != null@ && ...) {
          result.add(new CycleBoundTick(
            ..., @lastTick.getNumber@(), ...));
        }
      } else {
        if (@lastTick != null@ && ...) {
          result.add(new CycleBoundTick(
            ..., @lastTick.getNumber@(), ...));
        }
      }
    }
    lastTick = tick;
  }
}
\end{lstlisting}
		\caption{The reference program.}
		\label{fig:defect4jfix11}
	\end{subfigure}  
	\caption{A bug in Defects4J that \tool caught.}
    \label{fig:mbdexample}
\vspace{-8pt}
\end{figure*}

\subsection{Evaluating \tool on JBench}
\label{subsect:jbench}
Next, we evaluate the effectiveness of \tool in catching data races, {an} important class of programming errors in concurrency that is notoriously difficult to detect and fix. This is {a} specially challenging task considering \tool has never seen data races or even concurrent programs from training programs. Nevertheless, we think \tool can still perform well against data races considering many of the common errors that lead to data races are similar to those that appeared in our benchmark. For example, missing locks can be deemed as missing checks/statements; using non-atomic operations (\eg\,\texttt{i++} instead of \texttt{i.getAndIncrement()}) are often related to API misuse.

Similar to the previous experiment, we evaluate \tool in real scenarios of static bug finding. We use bugs provided in JBench~\cite{10.1145/3196398.3196451}, one of the most recent benchmark suites of data races in Java programs, as the groundtruth for this experiment. JBench contains 985 data races in total. We discard 204 extracted from research prototypes (\eg~\citet{Joshi2009},~\citet{Huang2014}, \etc) as they are quite different from concurrent programs in real-world applications.
We also exclude races triggered in library methods for which we can not generate the program dependence graph. We left with 657 for this evaluation. 
%



As the first step, we design the light-weight analysis to collect test programs that potentially have data races. A crucial task is to find shared variables. We follow the algorithm proposed by~\citet{kahlon2007fast}. Specifically, we perform a data-flow analysis to detect complete update sequences from \texttt{p} to \texttt{q} (\eg\,\texttt{p_1 = p;p_2 = p_1;\dots;q = p_k}) that are followed by the modification of a variable accessed via \texttt{q} (\eg\,\texttt{q.a = v;}), where \texttt{p} either points to a global variable or is passed as a parameter to an API function. A variable is declared as shared if (1) it is propagated from a complete update sequence; and (2) gets assigned by a local variable or an expression. After a shared variable is discovered, we regard all procedures that use the variable as the test program. Furthermore, we mark the write statement (\eg\,\texttt{q.a = v;}) and every statement that reads the variable in the update sequence (\eg any read to a variable among \texttt{p,p_1,p_2\dots,p_k,q}) to be the point at which a data race can occur. Regarding the graph format of test programs, we first represent each set with its interprocedural PDG. Then, we link all the interprocedural  PDGs via a data dependence  edge (from \texttt{p} to \texttt{q}) through the node of the shared variable on each interprocedural  PDG when the corresponding operations include at least one write. We left an example in the supplemental material (Section~\ref{app:edges}) for reviewers' perusal. For prediction interpreter, we introduce a simple change to our procedure for determining the feasibility of program paths that trigger data races. Because data races are triggered by concurrent sub-paths, both of which access the same shared variable. Thus, we solve the \textit{union} of the path constraints collected from both sub-paths to ensure their concurrent feasibility.

O2~\cite{Liu2021} is the state-of-the-art for static race detection, at the same time, it's a commercial tool that is not publicly accessible. For this reason, we compare \tool against RacerD~\cite{10.1145/3276514}, the then state-if-the-art static race detector. By default, RacerD does not perform a whole program analysis, therefore, it may miss code that \tool covers. To ensure RacerD having at least the same coverage as \tool, we manually annotate all classes (with @ThreadSafe annotations) in each project (included in JBench) that are covered by the test code \tool identified with shared variables.
We also include FICS as another baseline since FICS claims to be capable of detecting any type of bugs. FICS has the same configuration as the previous experiment which produces the best performance again on JBench. For \tool, we reuse the model trained for Defects4J evaluation. Same as before, we set the cut-off value $N$ to be 1 and 5 for this experiment.

\begin{table}[t]
	\captionsetup{skip=1pt}
	\centering
	\caption{Comparing \tool against RacerD and FICS on JBench. top-max sets the cut-off value $N$ to be the total number of bugs in a project. {\large\textbf{+}} is the number of bugs a tool finds but not others. When counting {\large\textbf{+}} for RacerD and FICS, we consider \tool in top-max which achieves the best performance. For clarity, we split the six projects in JBench into two groups. The first row shows the results of each tool on dbcps2, ftpserver and guava while the second row shows the results of each tool on the rest.}
	\begin{subtable}[h]{\textwidth}
			\centering
			\begin{tabular}{p{.1\textwidth}|p{.1\textwidth}|p{.03\textwidth}<{\centering} p{.03\textwidth}<{\centering} p{.03\textwidth}<{\centering} p{.03\textwidth}<{\centering} | p{.03\textwidth}<{\centering} p{.03\textwidth}<{\centering} p{.03\textwidth}<{\centering} p{.03\textwidth}<{\centering} | p{.03\textwidth}<{\centering} p{.03\textwidth}<{\centering} p{.03\textwidth}<{\centering} p{.03\textwidth}<{\centering}}
				\hline
				\multicolumn{2}{c|}{} & \multicolumn{4}{c|}{dbcps2} & \multicolumn{4}{c|}{ftpserver} & \multicolumn{4}{c}{guava} \\\cline{3-14}
				\multicolumn{2}{c|}{} & TP & FP & FN & {\large\textbf{+}} & TP & FP & FN & {\large\textbf{+}} & TP & FP & FN & {\large\textbf{+}} \\\hline
				\multicolumn{2}{c|}{RacerD} & 11 & 96 & 17  & 4 & 14 & 83  & 44 & 1 & 8 & 54 & 19 & 4 \\\hline
				\multicolumn{2}{c|}{FICS} & 0  & 52 & 28  & 0 & 5  & 115 & 53 & 0 & 1 & 4  & 26 & 0 \\\hline
				\multirow{3}{*}{\tool} 
				& top-1 & 1 & 0 & 27  & 0 & 1 & 0  & 57 & 0 & 1 & 0 & 26 & 0 \\\cline{2-14}
				& top-5 & 4 & 1 & 24  & 0 & 5 & 0  & 53 & 0 & 4 & 1 & 23 & 1 \\\cline{2-14}
				& top-max& 9& 19& 19  & 2 & 19& 39 & 39 & 6 & 8 & 19& 19 & 4\\\hline
			\end{tabular}
	\end{subtable}

	\vspace{12pt}

	\begin{subtable}[h]{\textwidth}
			\centering
			\begin{tabular}{p{.1\textwidth}|p{.1\textwidth}|p{.03\textwidth}<{\centering} p{.03\textwidth}<{\centering} p{.03\textwidth}<{\centering} p{.03\textwidth}<{\centering} | p{.03\textwidth}<{\centering} p{.03\textwidth}<{\centering} p{.03\textwidth}<{\centering} p{.03\textwidth}<{\centering} | p{.03\textwidth}<{\centering} p{.03\textwidth}<{\centering} p{.03\textwidth}<{\centering} p{.03\textwidth}<{\centering}}
				\hline
				\multicolumn{2}{c|}{} & \multicolumn{4}{c|}{log4j} & \multicolumn{4}{c|}{tomcat} & \multicolumn{4}{c}{zookeeper} \\\cline{3-14}
				\multicolumn{2}{c|}{} & TP & FP & FN & {\large\textbf{+}} & TP & FP & FN & {\large\textbf{+}} & TP & FP & FN & {\large\textbf{+}}  \\\hline
				\multicolumn{2}{c|}{RacerD} & 16 & 102 & 120 & 0 & 38 & 272 & 206 & 2 & 18 & 413 & 146 & 0 \\\hline
				\multicolumn{2}{c|}{FICS}   & 4  & 291 & 132 & 0 & 12 & 981 & 232 & 1 & 3  & 174 & 161 & 0 \\\hline
				\multirow{3}{*}{\tool}
				& top-1 & 1  & 0 & 135 & 0 & 1  & 0  & 243 & 0 & 1 & 0  & 163 & 0 \\\cline{2-14}
				& top-5 & 5  & 0 & 131 & 0 & 5  & 0  & 239 & 0 & 5 & 0  & 159 & 0 \\\cline{2-14}
				& top-max&44 &92 & 92  & 28 & 76 & 168& 168 & 40 & 61& 103 & 103& 43 \\\hline
			\end{tabular}
	 \end{subtable}
	 \label{tab:jbenchres}
\end{table}

\vspace{3pt}
\noindent
\textbf{\textit{Results.}}\, As shown in Table~\ref{tab:jbenchres}, \tool in top-1 and top-5 display almost a perfect precision, reporting very few false warnings. However, this is insignificant considering the number of data race that each project has far exceeds 5. For a more fair evaluation, we set $N$ to be the number of data races a project has in total, denoted by top-max in Table~\ref{tab:jbenchres}. \tool again comes on top. Compared to RacerD, which is clearly the better baseline, \tool on average has both significantly lower false positive rate (68\% vs. 89\%) and false negative rate (67\% vs. 78\%). Overall, we conclude that \tool is capable of detecting data races from concurrent programs, the type of code that is vastly different from what \tool has encountered during training. 

\subsection{A Comprehensive Investigation on the Results of \tool}
\label{subsec:deep}

In this section, we perform an in-depth analysis of \tool's results on Defects4J and JBench. Our goal is to gain an insight on why \tool is accurate in detecting bugs in some cases while lacks precision in others.

\vspace{3pt}
\noindent
\textbf{\textit{On the Strengths of \tool.}}\, We study 53 bugs \tool caught in top-5 prediction mode which is a good sample size for our analysis. To understand the reason why \tool predicts them to be buggy, for each bug, we gather the reference program under the relational embedding approach.\footnote{we explain what is a reference program for a given bug in the last paragraph of Section~\ref{subsec:def}.} First, we confirm the inconsistency indeed exists for each bug and its reference, subsequently, we manually annotate the part of the program in the bug and reference program that constitutes the inconsistency. Because every inconsistency can be expressed with no greater than 6 statements in the bug or reference program, we retrieve 6 statements assigned the highest global attention weight (Equation~\ref{equ:align2}) from both programs. After a thorough examination, we find that in 42 out of 53 bugs the annotated code are fully covered by the 6 highest attended statements in the bug and reference program, even for the few remaining bugs, the 6 highest attended statements miss at most 2 statements combined in the annotated code. This is a convincing piece of evidence that \tool has learned the right features that capture the semantic inconsistency between buggy and reference programs. Another interesting finding is that 17 reference programs used in this study exhibit program features (\eg unknown APIs, new language features like lambda expression, reflections, \etc) that never appear in the training set of \tool and rarely occur in other test programs collected in the same batch.\footnote{for each bug, we collect approximately 20 test programs on average. 17 reference programs refer to those that contain features that rarely occur in the other 19 test programs collected in the same batch.}
Under this circumstance, \tool could have been easily tricked to predict such reference programs as bugs, if it relied on syntactic variations. Instead, \tool is not confused. Even without the knowledge of those rare features, \tool can still determine whether or not they constitute a semantic inconsistency by analyzing how they are used in test programs.
This findings testify the precision of \tool in recognizing deep, semantic inconsistency beyond syntactic variations, which is the primary reason behind \tool's superior effectiveness to traditional static analysis tools like \eradicate, Infer or RacerD. The reason that \tool significantly outperforms FICS is also straightforward: FICS adopts an unsupervised approach; without a training phase like \tool's FICS do not have a chance to learn to pinpoint the kind of semantic inconsistency that leads to bugs. As a result, it seems to be easily confused by syntactic variations: flagging anything it sees as an inconsistency.

\vspace{3pt}
\noindent
\textbf{\textit{On the Limitations of \tool.}}\, We also attempt to understand why \tool fails to find bugs. We find that the main source of \tool's inaccuracy stems from its handling of exceedingly complicated programs. For example, \tool is shown to be ineffective against quite a few data races triggered by the interleaving of many threads, each of which yields a lengthy call chain spanning across dozens of procedures. The cause of the phenomenon is the well-known deficiency of graph neural networks in processing large, complex graphs. Despite for different reasons, this issue is in fact common to all tools used in our evaluation including both state-of-the-art static analysis tools and learning-based bug detectors. Nevertheless, it points to a worthwhile direction for future work on top of \met's contribution in this paper.
Another factor that contribute to \tool's imprecision is that bugs may not always be inconsistent with correct programs. For example, several bugs in array index out-of-bound involve specialized, heavy arithmetic operations for computing the bound of an array. Although the computations turn out to be erroneous, they are not necessarily inconsistent with any correct program. Thus, \tool did not flag them as buggy. This may be a limitation for \tool, however, considering bugs not showing inconsistency with correct programs are minorities, \tool's approach --- meta bug detection --- is still valid.

\subsection{On the Necessity of Key Components of \tool}
	\begin{wraptable}{r}{0.5\textwidth}
		\vspace{-14pt}
		\captionsetup{skip=1pt}
		\caption{Results of the ablation study.}
		\begin{tabular}{l|c}
		\hline
		Configuration & \# of Bugs Found \\\hline
		original & 53	\\\hline
		w/o global graph attention & 44 \\\hline
		w/o relational embedding & 46 \\\hline
		\end{tabular}
		\label{tab:abl}
	\end{wraptable} 


To confirm the necessity of each key component of \tool, we perform an ablation study to reveal their contributions. We compare \tool's performance on both Defects4J and JBench before and after the removal of a component. We use true positives as the only metric because false positives/negatives convey the same information given \tool generates the same number of reports before and after the ablation. For convenience, we sum the number of true positives \tool finds in Defects4J and JBench. Since the prediction mode is not a factor for this study, we present our results only in top-5. First, we drop the global attention mechanism from the GNN, and re-run experiments in the same procedure as before. As depicted in the second row of Table~\ref{tab:abl}, \tool catches 9 fewer bugs than before, a significant downgrade from its original configuration. This indicates the idea that the global graph attention realizes --- focusing on nodes that express the essential characteristics of graphs --- improves GNN's precision in meta bug detection.

Next, we forgo the relational embedding approach. As a result, \tool misses 7 bugs. In fact, only adding back the attention mechanism in the relational embedding (\ie without read operations) still has \tool catch 4 fewer bugs, not to mention, the complete relational embedding approach embeds every buggy program further away from the average program than both the partial (\ie without read operations in relational embedding) and the plain embedding approach (\ie without relational embedding at all). This strongly suggests that relational embedding helps to improve the precision of the learning approach,
furthermore, the read operations succeeded in amplifying the effect of attention in this relational embedding approach. Overall, our ablation study shows the components are all necessary for \tool to be effective.

\section{Related Work}
\label{sec:rela}

We survey related work from anomaly detection, formal methods- and learning-based bug finders.

\subsection{Anomaly Detection}
\citet{Engler502041} proposed to analyze bugs as deviant behavior, a seminal work in anomaly detection. DIDUCE~\cite{10.1145/581339.581377}, also based on anomaly detection, monitors the violation of invariants during the execution of programs. Bixie~\cite{mccarthy2015bixie, 10.1145/2491411.2491448} regards a code fragment to be buggy if it is not part of any normal terminating executions.
These works are by nature rule-based and focused on a few specific types of programming errors. Some bug detectors take a more general approach to identifying coding inconsistencies.
APIsan~\cite{197149} infers correct API usages in source code
through symbolic execution and semantic cross-checking. Similar to AntMiner~\cite{7886915} detects API usage inconsistencies via programming rules minded from the program dependence graph.~\citet{9519443} also detect API Misuses base on active learning. The work that is the closest to ours is FICS~\cite{263838}, which is also a bug-generic detector. However, FICS is built upon an unsupervised approach without a training phase. As a result, it is shown to be significantly less precise than \tool in our experiments. 

\subsection{Formal Methods-based Bug Finding}
Model checking has been an important technique for static bug finding. SPIN~\cite{Holzmann588521}, Java PathFinder~\cite{Visser2000JavaPathFinder}, and CBMC~\cite{Clarke2003, Clarke2004} are notable examples. SLAM~\cite{SLAM} and BLAST~\cite{BLAST} improve the scalability of the above tools with the idea of abstraction refinement. Data-flow analysis has also been applied extensively in this field (\eg IFDS~\cite{reps1995precise}, Saturn~\cite{Xie2005Saturn}, and CALYSTO~\cite{Babic1368118}). But these tools often suffer from scalability issues. Sparse value-flow analysis~\cite{Cherem1250789,Livshits940114,Sui2892235,pinpoint} mitigates this problem by tracking the flow of values sparsely through def-use chains or static single
assignment form. Infer\cite{calcagno2015moving,berdine2005smallfoot} finds bugs by automatically inferring separation logic assertions over statements. Compared to the static analysis tools, \tool benefits from a simpler design and higher utility, specifically, it can detect the types of bugs that are unobserved during training.

\subsection{Learning-based Bug Detection}
\citet{allamanis2018learning} develop a new program graph by incorporating the data flow and type hierarchies information into ASTs for predicting the name of a variable and detecting the misuse of variable.~\citet{vasic2018neural} present multi-headed pointer networks for detecting the same variable misuse bug.~\citet{Hellendoorn2020Global} further improve the joint model~\cite{vasic2018neural} by combining sequence model (\eg, RNN, transformer) with a structure model like GNN. DeepBugs~\cite{pradel2018deepbugs} presents a learning approach to name-based bug detection.~\citet{Wang3428205} 
develop an interval-based graph abstraction method to improve the scalability of GNNs. \met is fundamentally different from all the works above which are dedicated to learning bug patterns.


\vspace{4pt}
\section{Conclusion}
\label{sec:conc}

In this paper, we propose \textit{meta bug detection}, a fundamentally different concept from the methodology of existing learning-based bug detection. Our insight is to learn how bugs are inconsistent with correct programs independently of the type of bugs. Built on top of this insight, \met (1) requires substantially less training data than the existing bug detectors, (2) is capable of predicting the type of bugs that are totally unobserved during training, and (3) can explain its prediction without any external interpretability methods.
We realize \met into a tool, \tool, and extensively evaluated it. On well-established datasets like Defects4J and JBench, 
results show \tool is effective, catching a large number of bugs, in particular it significantly outperforms noteworthy static analysis tools like Facebook Infer and RacerD, and learning-based anomaly or bug detectors like FICS and GINN.

\clearpage
\bibliography{reference}

\clearpage
\appendix

\section{Test Data Collector For Common Types of Bugs}
\label{append:rules}
To collect test data for array index out-of-bound, we first find out the array operations $O$ including array read or write at a program point $l$. For each $o\in O$, we then use the array operation of array $arr$ and index $idx$ at $l$ as the criterion to compute a program slice which include the entire data dependence of $arr$ and $idx$. Finally, we regard all such slices to be the test code for array index out-of-bound.  

For number format exception, we first find out the operations $O$ that parse strings $s$ to numbers. We then use each $o\in O$ as the criterion to compute a program slice which include the entire data dependence of $s$. Finally, we regard all such slices to be the test code for number format exception.

\clearpage

\section{An Example for \textsf{MBD}'s Correctness Definition}
\label{append:explanationexample}
\begin{figure}[h!]
	\centering
	\captionsetup{skip=4pt}
	
	\parbox{.5\linewidth}{
		\lstset{style=mystyle}
\begin{lstlisting}[morekeywords={Object, String},
		% numbers=left, stepnumber=1, xleftmargin=2em, frame=single, framexleftmargin=1.5em, 
		basicstyle=\scriptsize\ttfamily\bfseries]
  void foo(int x, int y) {
1:  x = 5;
2:  y = 6;
3:  log("value of x,y:", x, y);
4:  y = Math.abs(x) /* y is assign the 
                       absolute value of x */
    if (y >= 0) 
    {
5:    str = null;
6:    str.length();
    }
  }
\end{lstlisting}
	}
	\caption{An example for Definition 3.3.}
	\label{fig:explanationexample} 
\end{figure}

For the example in Figure~\ref{fig:explanationexample}, $T$ is $1 \rightarrow 2 \rightarrow 3 \rightarrow 4 \rightarrow 5 \rightarrow 6$, and $T'$ is $1 \rightarrow 4 \rightarrow 5 \rightarrow 6$, however, the correct trace $\bar{T}$ should be $4 \rightarrow 5 \rightarrow 6$ because the value of $y$ will always be non-negative. In fact, computing the minimal unsatisfiable cores of the symbolic trace will be useful in finding $\bar{T}$, specifically, \texttt{y = Math.abs(x)} $\land$ \texttt{y > 0} $\land$ \texttt{str = null} $\land$ \texttt{assert(str != null)}\footnote{we represent the semantic constraint expressed by \texttt{str.length())} with \texttt{assert(str != null)}.} is the minimal unsatisfiable cores of the entire trace. But it is out of the scope of our paper which focus on meta bug detection instead of error trace explanation. 

\subsection{Definition of Subsequence}
A subsequence of <$a$> is a sequence <$b$> defined by $b_k=a_{n_k}$, where $n_1<n_2<...$ is an increasing sequence of indices~\cite{john1997mathematical}. For example, if $a_n = 2n - 1$ and $n_k = k^2$, then $b_k = 2k^2 - 1$~\cite{john1997mathematical}.

\begin{table}[H]
	\centering
	\begin{tabular}{c|ccccccccc}
		$n$ & 1 & 2 & 3 & 4 & 5 & 6 & 7 & 8 & 9 \\
		$a_n$ & 1 & 3 & 5 & 7 & 9 & 11 & 13 & 15 & 17 \\
		$k$ & 1 &  &  & 2 &   &  &  &  & 3 \\
		$b_k$ & 1 &  &  & 7 &   &  &  &  & 17 \\
	\end{tabular}
\end{table}


\clearpage

\section{An Interprocedural Program Dependence Graph}
\label{append:interprocedural}

We show the entire interprocedural program dependence graph for the program in Figure~\ref{fig:intercode} of the main paper.

\begin{figure}[h]
	\centering
	\captionsetup{skip=4pt}
	\includegraphics[height=12cm]{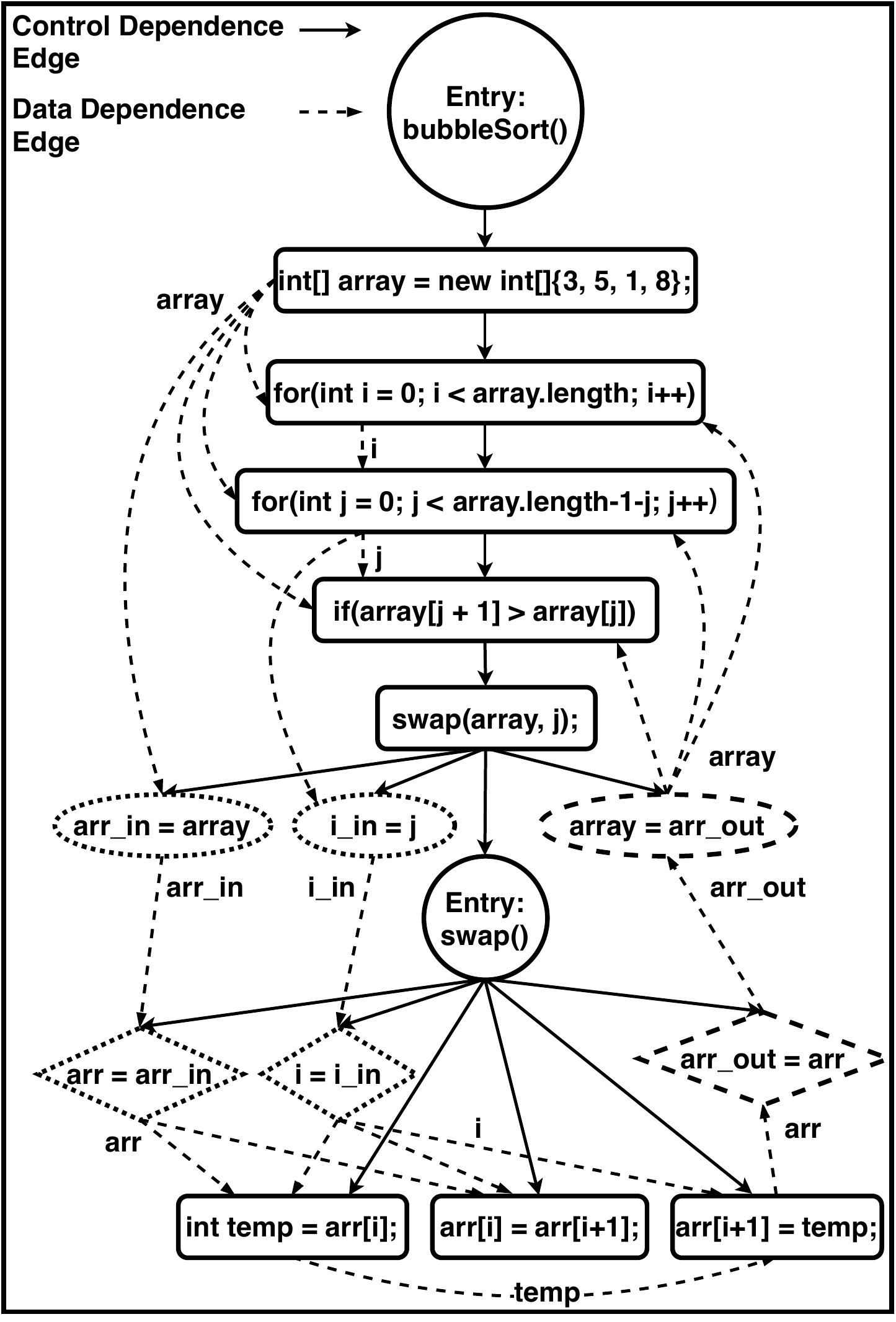}
	\caption{The entire interprocedural program dependence graph for the program in Figure~\ref{fig:intercode} of the main paper. We omit AST nodes and edges, as well as execution order edges for brevity.}
	\label{fig:interAll}	
\end{figure}

\clearpage

\section{Defects4J Bugs Detected by Different Tools}
\label{append:onebyone}

\begin{table*}[h!]
	\captionsetup{skip=1pt}
	\caption{Null pointer dereference in Defects4J detected by Infer, \eradicate, FICS and \tool.}
	\begin{tabular}{l|p{.08\textwidth}<{\centering}|p{.08\textwidth}<{\centering}|p{.08\textwidth}<{\centering}|p{.13\textwidth}<{\centering}|p{.13\textwidth}<{\centering}}
		\hline
		Proj. & Infer & \eradicate & FICS & \tool (Top-1) & \tool (Top-5) \\\hline
		Chart-2 & & & & & \\\hline
		Chart-4 & & & & & \\\hline
		Chart-14 & & \ding{52} & & & \\\hline
		Chart-16 & & \ding{52} & & & \\\hline
		Cli-5 & \ding{52} & \ding{52} & & & \\\hline
		Cli-30 & & & & & \\\hline
		Codec-5 & & & & & \\\hline
		Codec-13 & & \ding{52} & & & \\\hline
		Codec-17 & & \ding{52} & & \ding{52} & \ding{52} \\\hline
		Collections-25 & & \ding{52} & & & \\\hline
		Collections-28 & & & & & \\\hline
		Csv-4 & & & & & \ding{52} \\\hline
		Csv-9 & & & \ding{52} & & \\\hline
		Csv-11 & & & & \ding{52} & \ding{52} \\\hline
		JacksonCore-8 & & & & \ding{52} & \ding{52} \\\hline
		JacksonDatabind-3 & & & & & \ding{52} \\\hline
		JacksonDatabind-13 & & & & & \\\hline
		JacksonDatabind-36 & & & & & \\\hline
		JacksonDatabind-80 & & \ding{52} & & & \\\hline
		JacksonDatabind-93 & & & & & \\\hline
		JacksonDatabind-95 & & \ding{52} & & & \\\hline
		JacksonDatabind-107 & & \ding{52} & & \ding{52} & \ding{52} \\\hline
		Jsoup-8 & \ding{52} & \ding{52} & & & \\\hline
		Jsoup-22 & & & & & \\\hline
		Jsoup-26 & & & & & \\\hline
		Jsoup-66 & & & & & \\\hline
		Jsoup-89 & & & \ding{52} & & \\\hline
		Lang-20 & & & & & \\\hline
		Lang-33 & & & \ding{52} & \ding{52} & \ding{52} \\\hline
		Lang-39 & & & & & \ding{52} \\\hline
		Lang-47 & & & & & \\\hline
		Lang-57 & & & & & \\\hline
		Math-4 & \ding{52} & \ding{52} & & & \\\hline
		Math-70 & & & & & \\\hline
		Math-79 & & & & & \ding{52} \\\hline
		Mockito-18 & & & & & \\\hline
		Mockito-35 & & & & \ding{52} & \ding{52} \\\hline
		Mockito-36 & & & & & \\\hline
		Mockito-38 & & & & & \ding{52} \\\hline
		\multicolumn{6}{l}{\small \ding{52} indicates the bug was correctly detected by the tool.}
	\end{tabular}
	\label{tab:npddetail}
\end{table*}

\begin{table*}[h!]
	\captionsetup{skip=1pt}
	\caption{Array index out-of-bound in Defects4J detected by FICS and Infrared.}
	\begin{tabular}{p{.17\textwidth}|p{.15\textwidth}<{\centering}|p{.15\textwidth}<{\centering}|p{.15\textwidth}<{\centering}}
		\hline
		Proj. & FICS & \tool (Top-1) & \tool (Top-5) \\\hline
		Chart-5 & & & \ding{52} \\\hline
		Chart-18 & & \ding{52} & \ding{52} \\\hline
		Chart-22 & & & \\\hline
		Cli-32 & & & \\\hline
		Codec-8 & \ding{52} & & \\\hline
		Codec-18 & & \ding{52} & \ding{52} \\\hline
		Compress-21 & & \ding{52} & \ding{52} \\\hline
		Compress-37 & & & \\\hline
		JacksonCore-10 & \ding{52} & & \\\hline
		JacksonCore-11 & \ding{52} & & \\\hline
		JacksonCore-19 & & & \\\hline
		JacksonCore-25 & & & \\\hline
		Jsoup-5 & & & \\\hline
		Jsoup-34 & & & \\\hline
		Jsoup-72 & & & \\\hline
		Jsoup-80 & & & \\\hline
		Jsoup-86 & & & \\\hline
		Jsoup-90 & & & \\\hline
		Lang-6 & & & \ding{52} \\\hline
		Lang-12 & \ding{52} & & \\\hline
		Lang-19 & & \ding{52} & \ding{52} \\\hline
		Lang-27 & & & \\\hline
		Lang-44 & \ding{52} & & \\\hline
		Lang-45 & & & \\\hline
		Lang-51 & & & \\\hline
		Lang-59 & & \ding{52} & \ding{52} \\\hline
		Lang-61 & & & \\\hline
		Math-3 & & & \\\hline
		Math-81 & & & \ding{52} \\\hline
		Math-98 & & & \\\hline
		Math-100 & & \ding{52} & \ding{52} \\\hline
		Math-101 & \ding{52} & & \\\hline
		Mockito-3 & & & \\\hline
		Mockito-34 & & & \ding{52} \\\hline
		\multicolumn{4}{l}{\small \ding{52} indicates the bug was correctly detected by the tool.}
	\end{tabular}
	\label{tab:aiodetail}
\end{table*}

\begin{table*}[h!]
	\captionsetup{skip=1pt}
	\caption{Number format exception in Defects4J detected by FICS and Infrared.}
	\begin{tabular}{p{.17\textwidth}|p{.15\textwidth}<{\centering}|p{.15\textwidth}<{\centering}|p{.15\textwidth}<{\centering}}
		\hline
		Proj. & FICS & \tool (Top-1) & \tool(Top-5) \\\hline
		Compress-32 & & & \ding{52} \\\hline
		JacksonCore-5 & \ding{52} & \ding{52} & \ding{52} \\\hline
		Lang-1 & & & \\\hline
		Lang-16 & & & \ding{52} \\\hline
		Lang-36 & & \ding{52} & \ding{52} \\\hline
		Lang-58 & & & \\\hline
		\multicolumn{4}{l}{\small \ding{52} indicates the bug was correctly detected by the tool.}
	\end{tabular}
	\label{tab:nfedetail}
\end{table*}

\clearpage

\section{Comparing \eradicate and Infer's Results in Our Evaluation against~\citet{Tomassi2021}} \label{append:manual}

%

We describe how we confirm the results of baseline methods (\eg \eradicate and Infer) match those of~\citet{Tomassi2021}. After a thorough review of their detailed experiment results\footnote{released at \url{https://github.com/ucd-plse/Static-Bug-Detectors-ASE-Artifact/tree/main/data}}, we summarize our findings below. First, for each bug in their dataset, they run \eradicate and Infer twice, one on the buggy version and the other on the fixed version\footnote{such that the presence/absence of a bug report on the buggy/fixed version of the project points to the bug.}, and they count the same warning twice when reporting the total number of alarms. Third, bugs are always analyzed separately in~\citet{Tomassi2021} even if they are from the same project. As a result, a warning occurs as many times as the project is analyzed. Since~\citet{Tomassi2021} did not remove duplicate alarms, the same report produced by \eradicate and Infer can be countered multiple times. Finally,~\citet{Tomassi2021} also count the warnings \eradicate and Infer report on code written for testing purposes. As we explained in the main paper, we do not consider bugs in test code for our evaluation, in the same way, we also ignore all of \eradicate and Infer's reports on test code. Taking into account all factors above, we confirm their results indeed match ours.

%
%
%
%
%

\clearpage

\section{A Data Race Example}
\label{app:edges}

\begin{figure}[h!]
	\centering
	\captionsetup{skip=4pt}
	
	\parbox{.5\linewidth}{
		\lstset{style=mystyle}
\begin{lstlisting}[morekeywords={Object, String}, numbers=left, stepnumber=1, xleftmargin=2em, frame=single, framexleftmargin=1.5em, basicstyle=\scriptsize\ttfamily\bfseries]
class Main {
  ChildThread childThread;
  void execute() {
    childThread = new ChildThread(this);
    childThread.start();
    ...
    synchronized (this) {
      if (childThread != null) {
        childThread.interrupt();
      }
    }
  }
}

class ChildThread extends Thread {
  Main main;
  ChildThread(Main main) {
    this.main = main;
  }
  void run() {
    ...
    main.childThread = null;
  }
}
\end{lstlisting}
	}
	\caption{A program with data race.}
	\label{fig:racesource} 
\end{figure}

\begin{figure}[h!]
	\centering
	\captionsetup{skip=4pt}
	
	\includegraphics[height=6cm]{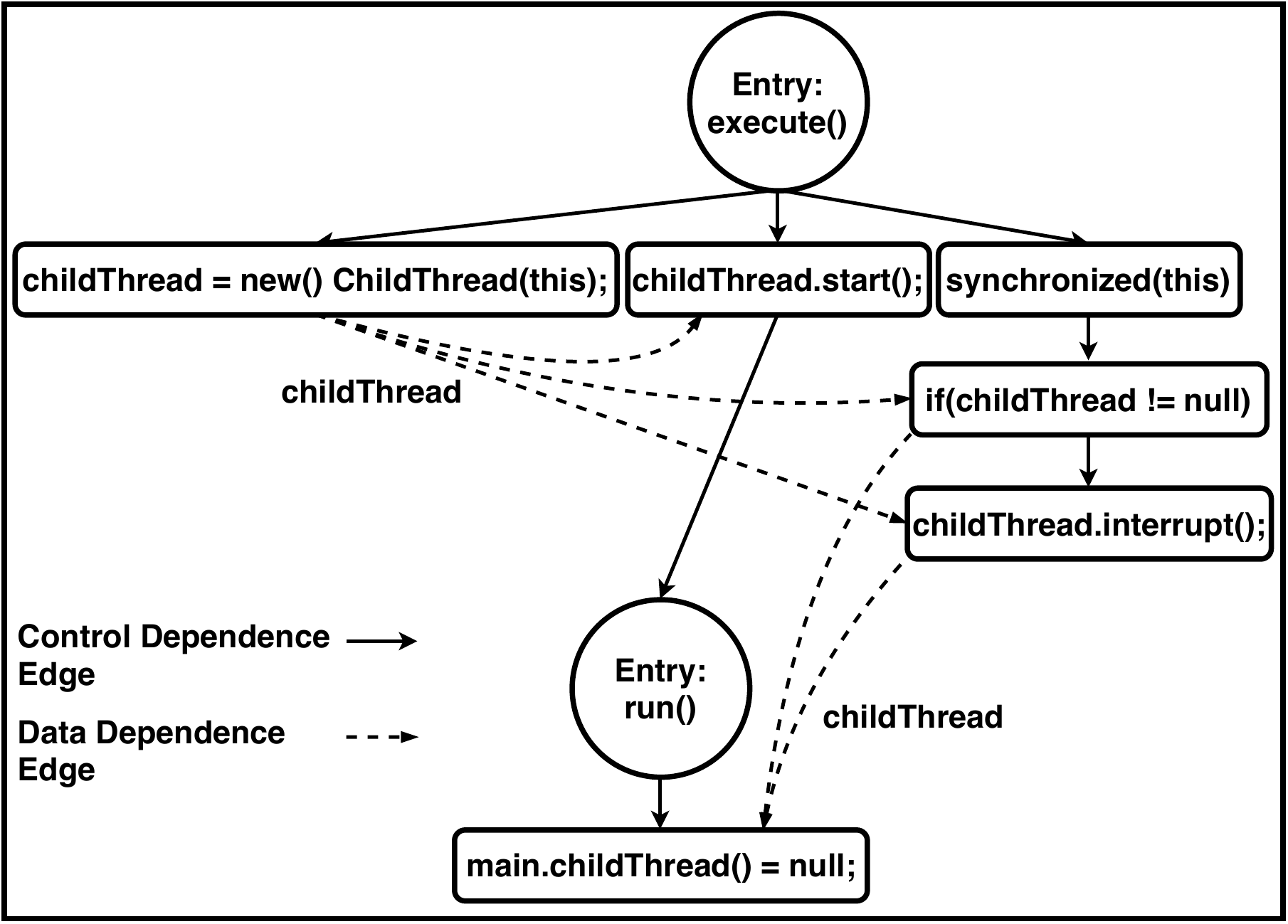}
	\caption{The graph representation of the program in Figure~\ref{fig:racesource}. We add two data dependence edges between the PDG of \texttt{Main::execute} and \texttt{ChildThread::run} to denote uses of the shared variable.}
	\label{fig:racerep}
\end{figure}

We use the example from ~\citet{10.1145/966049.781528} to illustrate our approach. In Figure~\ref{fig:racesource}, the main thread starts a new thread called \texttt{ChildThread} (Line 4) and then tries to terminate it (Line 9). Because  the only lock at line 7 cannot prevent the simultaneous access to a shared variable \texttt{childThread}. In particular, when  \texttt{main.childThread = null} (Line 22) is executed between \texttt{if(childThread != null)} (Line 8) and  \texttt{childThread.interrupt()} (Line 9), a null pointer dereference will be thrown.

To analyze this program, we first found a shared variable \texttt{this} in method \texttt{Main::execute} according to (1) a complete update sequence: \texttt{main = this;} (the parameter passed to the constructor of \texttt{ChildThread}), and \texttt{this.main = main;} (inside of the constructor of \texttt{ChildThread}); and (2) a subsequent modification of the field variable \texttt{childThread} accessed via \texttt{this.main} (inside of \texttt{ChildThread::run}).\footnote{The variable \texttt{this} in \texttt{Main::execute} and \texttt{this.main} in \texttt{ChildThread::run} can be regarded as \texttt{p} and \texttt{q} in the algorithm we explained in the main paper.} Then, we identify \texttt{Main::execute} and \texttt{ChildThread::run} as the methods that use this variable. Therefore, they become a test program. For the graph representation, we first construct the interprocedure PDG (each containing only one procedure in this specific case) for each method, and then we connect the use points of the shared variable in each interprocedure PDG as Figure ~\ref{fig:racerep} shows.

\end{document}